\documentstyle[aps,prl,preprint,floats,epsfig]{revtex}

\begin{document}

\tighten

\preprint{\tighten\vbox{\hbox{\bf CLNS 00/1664}
                        \hbox{\bf CLEO 00-3}
                        \hbox{\date{\today}}}}

\title{\boldmath Measurement of ${\cal B}(\Lambda_c^+\to pK^-\pi^+)$}

\author{CLEO Collaboration}
\date{\today}

\maketitle
\tighten

\begin{abstract}
The $\Lambda_c^+\to pK^-\pi^+$ yield has been measured in a 
sample of two-jet
continuum events containing both a charm tag (``${\overline D}$'') as
well as an antiproton ($e^+e^-\to \overline{D}{\overline p}$X), 
with the antiproton in the
hemisphere opposite the $\overline{D}$ (measurement
of charge conjugate modes is implicit 
throughout). Under the hypothesis that such
selection criteria 
tag $e^+e^-\to \overline{D}{\overline p}{\Lambda_c^+}$X events, the
$\Lambda_c^+\to pK^-\pi^+$ branching fraction can be determined by 
measuring the $pK^-\pi^+$ yield in the same hemisphere as the 
antiprotons in
our $\overline{D}{\overline p}$X sample. Three types of 
${\overline D}$
charm tags are used -
$\pi_{\rm soft}^-$ (from $D^{*-} \to \overline{D}^0\pi^-_{\rm soft}$), 
electrons (from 
$\overline{D} \to Xe^-\nu$), and fully
reconstructed $\overline{D}^0 \to K^+\pi^-$ or 
$D^- \to K^+\pi^-\pi^-$ or $D_{\rm s}^- \to \phi \pi^-$.
Combining our results obtained from the three independent 
charm tags,
we obtain ${\cal B}(\Lambda_c^+\to pK^-\pi^+)$ = 
(5.0 $\pm$ 0.5 $\pm$ 1.2)\%. \\

\end{abstract}


\newpage
\renewcommand{\thefootnote}{\fnsymbol{footnote}}

\begin{center}
D.~E.~Jaffe,$^{1}$ G.~Masek,$^{1}$ H.~P.~Paar,$^{1}$
E.~M.~Potter,$^{1}$ S.~Prell,$^{1}$ V.~Sharma,$^{1}$
D.~M.~Asner,$^{2}$ A.~Eppich,$^{2}$ J.~Gronberg,$^{2}$
T.~S.~Hill,$^{2}$
% ...
\ and \
\end{center}
 
\small
\begin{center}
$^{1}${University of California, San Diego, La Jolla, California 92093}\\
$^{2}${University of California, Santa Barbara, California 93106}\\
% ...
\end{center}

\newpage

\section{Introduction}
Of the four fundamental normalization branching fractions of charmed
hadrons (${\cal B}(D^0\to K^-\pi^+)$, 
${\cal B}(D^+\to K^-\pi^+\pi^+)$, 
${\cal B}(D_{\rm s}^+\to\phi\pi^+)$,
and ${\cal B}(\Lambda_c^+\to pK^-\pi^+)$),\footnote{Charge conjugate 
modes are implicit.}
the $\Lambda_c^+\to pK^-\pi^+$ branching fraction is the least well-known, and
presently the most controversial.
There have been two basic methods used to
estimate this branching fraction. The first uses, as input, the
ratio of efficiency-corrected yields: 
${{\cal B}(\Lambda_c^+\to\Lambda Xl\nu)\over{\cal B}(\Lambda_c^+\to 
pK^-\pi^+)}$\cite{ARGUS-lamcXlnu,CLEO-lamcXlnu} and the well-measured
$\Lambda_c^+$ lifetime. 
One can deduce a total
semileptonic branching fraction for $\Lambda_c^+$ decays
$${\cal B}(\Lambda_c^+\to Xl\nu)=
{\Gamma(\Lambda_c^+\to Xl\nu)\over\Gamma^{\rm tot}(\Lambda_c^+)},$$
assuming that the total semileptonic width is
the same in 
$\Lambda_c^+$ decays as in $D_{\rm s}^+\to Xl\nu$, $D^0\to Xl\nu$, and
$D^+\to Xl\nu$ (the approximate
equality of the semileptonic widths for all the charmed
mesons lends credence to this assumption, although mass and
phase space effects in semileptonic decays may be
significant\cite{dissenting-theorists}), and
assuming
${{\cal B}(\Lambda_c^+\to\Lambda Xl\nu)\over{\cal B}(\Lambda_c^+\to 
Xl\nu)}\approx$1.0\cite{CLEO93,ARGUS-Btobaryons,Isi98} 
(i.e., ${{\cal B}(\Lambda_c^+\to N{\overline K}Xl\nu)
\over{\cal B}(\Lambda_c^+\to Xl\nu)}\to 0$). 
Under these assumptions, one can 
estimate the absolute branching fraction for
$\Lambda_c^+\to\Lambda Xl\nu$, and, correspondingly, the absolute
branching fraction for $\Lambda_c^+\to pK^-\pi^+$ from the 
measured 
${{\cal B}(\Lambda_c^+\to\Lambda Xl\nu)
\over{\cal B}(\Lambda_c^+\to pK^-\pi^+)}$ yields. 
Such a procedure
yields values in
the range
of ${\cal B}(\Lambda_c^+\to pK^-\pi^+)\sim$6-8\%\cite{Isi98}. 

In the second approach, one uses
the fact that baryon number must be conserved in $B$-decay and that 
${\cal B}(b\to c)\approx$1.0. Under the assumption that baryon production in
$B$-decay occurs
through ${\overline B}\to\Lambda_c^+{\overline p}W$, the 
observed ${\overline B}\to{\overline p}X$ events provide an unbiased sample of
${\overline B}\to\Lambda_c^+X$.
Measurements of the $\Lambda_c^+\to pK^-\pi^+$
yield in such events therefore allow a determination of the
absolute $\Lambda_c^+\to pK^-\pi^+$ 
branching fraction\cite{CLEO93,ARGUS-Btobaryons}.\footnote{Unfortunately, 
a more recent
study of flavor-tagged baryon production in $B$-decay indicates
that diagrams other than  
${\overline B}\to\Lambda_c^+{\overline p}W$ may contribute substantially
to $\Lambda_c^+$, ${\overline\Lambda_c^-}$, 
and $p/{\overline p}$ production in 
$B$-decay\cite{flavor-tags}.}
The Particle Data Group uses a combination of this technique and
$D$ and $\Lambda_c^+$
charm semileptonic measurements to
estimate ${\cal B}(\Lambda_c^+\to pK^-\pi^+)=(5.0\pm1.3)$\%\cite{PDG98}. 

In this measurement, we employ a new technique to determine
${\cal B}(\Lambda_c^+\to pK^-\pi^+)$ using 
$e^+e^-$ annihilation
continuum events.
We select a sample of $e^+e^- \to c{\overline c}$ events in which a
$\Lambda_c^+$ is expected to be present by requiring: (i) a charm tag
consisting of either a high momentum electron, a $\pi^-_{\rm soft}$
(from $D^{*-}\to {\overline D^0}\pi^-_{\rm soft}$), or a fully
reconstructed ${\overline D}$-meson candidate and (ii) an
opposite hemisphere baryon tag
consisting of an antiproton. The presence of a $\Lambda_c^+$ is
inferred, to compensate baryon number and charm.
According to Monte Carlo
simulations, the antiproton in ${\overline D}{\overline p}\Lambda_c^+$
events is as likely to have its momentum in the same hemisphere 
as the ${\overline D}$ as in the
hemisphere opposite it.  However, estimation of the non-$\Lambda_c^+$
background in our ${\overline D}{\overline p}(\Lambda_c^+)$ 
sample is more
reliable if we require the antiproton to be in the hemisphere opposite
the charm tag.  We therefore focus on the sample in which the antiproton is
in the hemisphere opposite the charm tag (``O(${\overline p}|{\overline
D}$)'' events, with parentheses designating opposite hemisphere
correlations).\footnote{The same-hemisphere ${\overline p}{\overline D}$ 
sample, designated
with brackets as 
``S[${\overline p}{\overline D}$]'' is discussed later as a 
cross-check.} Topologically, these events can be schematically
depicted as: 
\begin{center}\[\begin{array}{cccc}
 & c &  \overline{c} & \\
 (\Lambda_c^+)\longleftarrow & & &  \longrightarrow {\overline D} \\
{\overline p}\longleftarrow & & & \\
\end{array}\]\end{center}

The yield of $\Lambda_c^+ \to pK^-\pi^+$ decays in this (${\overline
p}|{\overline D}$) sample will allow us, after all the appropriate
corrections, to determine the branching fraction:
\[{\cal B}(\Lambda_c^+\to pK^-\pi^+) =
{N(|[{\overline p}\Lambda_c^+]|{\overline D})\over N({\overline p}|{\overline
D})}.\] 

Our analysis comprises
two techniques -- in one, we construct a three-particle correlation to
determine the $\Lambda_c^+\to pK^-\pi^+$ branching fraction, 
and in the second,
a two-particle correlation is sufficient to 
infer ${\cal B}(\Lambda_c^+\to pK^-\pi^+)$.
In the triple correlation analysis, we take the ratio of the number of
times that three particles (the $\Lambda_c^+$, antiproton, and our 
charm tag)
are found in the same event relative to the number of times that only
the antiproton and the charm tag are found. For the second technique, 
only a double correlation between the
reconstructed $\Lambda_c^+$ and the antiproton tag constitutes the numerator
of our ratio; the recoiling charm tag is assumed.


\section{Apparatus and Event Selection}
\label{sec:event_selection}

This analysis was performed using the CLEO~II detector operating at the
Cornell Electron Storage Ring (CESR) at center-of-mass energies $\sqrt{s}$
= 10.52--10.58 GeV.  
The CLEO~II detector is a general purpose solenoidal magnet
spectrometer and calorimeter designed to trigger efficiently on two-photon,
tau-pair, and hadronic events\cite{kubota92}.  
Measurements of charged particle momenta are made with
three nested coaxial drift chambers consisting of 6, 10, and 51 layers,
respectively.  These chambers fill the volume from $r$=3 cm to $r$=1 m, with
$r$ being the radial coordinate relative to the beam (${\hat z}$) axis. 
This system is very efficient ($\epsilon\ge$98\%) 
for detecting tracks that have transverse momenta ($p_T$)
relative to the
beam axis greater than 200 MeV/$c$, and that are contained within the good
fiducial volume of the drift chamber ($|\cos\theta|<$0.94, with $\theta$
defined as the polar angle relative to the beam axis). 
This system achieves a momentum resolution of $(\delta p/p)^2 =
(0.0015p)^2 + (0.005)^2$ ($p$ is the momentum, measured in GeV/$c$). 
Pulse height measurements in the main drift chamber provide specific
ionization resolution
of 5.5\% for Bhabha events, giving good $K/\pi$ separation for tracks with
momenta up to 700 MeV/$c$ and separation of order 2$\sigma$ in the relativistic
rise region above 2 GeV/$c$. 
Outside the central tracking chambers are plastic
scintillation counters, which are used as a fast element in the trigger system
and also provide particle identification information from time-of-flight
measurements.  

Beyond the time-of-flight system is the electromagnetic calorimeter,
consisting of 7800 thallium-doped CsI crystals.  The central ``barrel'' region
of the calorimeter covers about 75\% of the solid angle and has an energy
resolution which is empirically found to follow
\begin{equation}
\frac{ \sigma_{\rm E}}{E}(\%) = \frac{0.35}{E^{0.75}} + 1.9 - 0.1E;
                                \label{eq:resolution1}
\end{equation}
$E$ is the shower energy in GeV. This parameterization includes
effects such as noise, and translates to an
energy resolution of about 4\% at 100 MeV and 1.2\% at 5 GeV. Two end-cap
regions of the crystal calorimeter extend solid angle coverage to about 95\%
of $4\pi$, although energy resolution is not as good as that of the
barrel region. 
The tracking system, time-of-flight counters, and calorimeter
are all contained 
within a superconducting coil operated at 1.5 Tesla. 
Flux return and tracking
chambers used for muon detection are located immediately outside the coil and 
in the two end-cap regions.

The event
sample used for this measurement is comprised of 3.1 $fb^{-1}$ of data
collected at the $\Upsilon$(4S) resonance and 1.6 $fb^{-1}$ of data 
collected about 60 MeV below the $\Upsilon$(4S) resonance. Approximately
$5\times 10^6$ continuum $c{\overline c}$ events are included in this sample.

\subsection{Event Selection Criteria}
In order to suppress background and enrich the
hadronic fraction of our event sample, we impose several event requirements. 
Candidate events must have: (1) at least four
detected, good quality, charged tracks; (2) an event vertex consistent with
the known $e^+e^-$ interaction point; (3) a total measured visible 
event energy, defined as the sum of the measured energy carried by
charged tracks plus the measured energy carried by neutral particles
($E_{\rm vis} = E_{\rm chrg} + E_{\rm neutral}$) 
greater than 110\% of the single
beam energy, $E_{\rm vis}$ $>$ 1.1 $\cdot$ $E_{\rm beam}$.  
In addition, when using
an electron to tag a $c{\overline c}$ event 
we require that either the beam energy
$E_{\rm beam}$ 
be less than 5.275 GeV (below the $\Upsilon$(4S)$\to B{\overline B}$
threshold) or that the event be well collimated.
Specifically, the ratio of Fox-Wolfram event shape parameters $H2/H0$ 
can be used to quantify the ``jettiness'' of an event\cite{FoxWolf} 
-- for a perfectly
spherical flow of event energy, this ratio equals 0; for a perfectly 
jetty event, this ratio equals 1.0. For our electron tags,
we require this ratio to be greater than 0.35.
This final requirement helps remove contamination from 
semileptonic $B$-decays in
$B{\overline B}$ events.
(The correlation between the soft pion momentum vector and the 
thrust 
axis is absent in $B\overline{B}$ events, 
therefore $B\overline{B}$ events do not contribute to our soft pion-tagged 
event sample.)

\section{Tag Identification}
\label{sec:tag_id}

\subsection{Charm Tags}
For our analysis, we
select continuum hadronic events which, in addition to an antiproton,
contain either a high momentum electron
(from ${\overline D}\to Xe\nu$), a $\pi^-_{\rm soft}$
(from $D^{*-}\to {\overline D^0}\pi^-_{\rm soft}$), 
or a fully reconstructed
${\overline D}$-meson candidate as a charm tag (``${\overline D}$'') of
$e^+e^-\to c{\overline c}$ events. 
Since the different tags have different 
systematic uncertainties and
procedures associated with them,
we now discuss separately the various tags employed
in this measurement, beginning with our electron charm tags.

\subsubsection{Electron Tags}
To suppress background from fake electrons, as well as 
true electrons not necessarily
associated with charm decays in $e^+e^- \to$ $c{\overline c}$ events, 
we require that our electron-tag
candidates satisfy the following criteria:

(a) The electron must pass a strict ``probability of electron''
identification criterion. This identification likelihood 
combines measurements of 
a given track's specific ionization deposition in the central drift chamber
with the ratio of the energy of the associated calorimeter shower to the
charged track's momentum\cite{ELECTRON-ID-REF}. 
True electrons have shower energies approximately
equal to their drift chamber momenta;
hadrons tend to be minimum ionizing and have considerably
smaller values of shower energy relative to their measured momenta.
We require that the logarithm of the
ratio of a charged track's
electron probability relative to the probability that the charged track
is a hadron be greater than 7.0. In the good fiducial volume of the CLEO
detector ($|\cos\theta|<$0.7, where $\theta$ is the track's polar
angle measured relative to the $e^+e^-$ beam axis), the efficiency of
this requirement is $>$90\% in our momentum interval of interest; the
likelihood of a non-electron faking an electron is less than 1\%. The
total electron
fake fraction is thus
the product of the fake rate per track times the typical
charged track multiplicity and is therefore not large ($\leq$10\%).

(b) The momentum of the electron must be greater than 1 GeV/$c$.  This 
criterion
helps eliminate fake electrons due to kaon and
pion tracks and also suppresses electrons from
photon conversions
($\gamma \to e^+ e^-$) and $\pi^0$ Dalitz
decays ($\pi^0 \to \gamma e^+ e^-$).

(c) The electron must have an impact parameter 
(``DOCA'', or distance-of-closest-approach)
relative to the primary event vertex
of less than 4 mm along the radial coordinate and no more than 2 cm along
the beam axis. This provides additional suppression of electrons resulting
from photon conversions.

\subsubsection{Soft-pion Tags}

Our soft-pion tag candidates must pass the following restrictions:

(a) The pion must have an impact parameter relative to the event vertex of
less than 5 mm along the radial coordinate and no more than 5 cm along
the beam axis.

(b) The pion must pass a 99\% probability criterion for pion identification,
based on the associated specific ionization collected in the drift chamber.

(c) The pion's momentum must lie between 0.15 GeV/$c$ and 0.40 GeV/$c$.

\begin{figure}
\begin{picture}(200,250)
\includegraphics{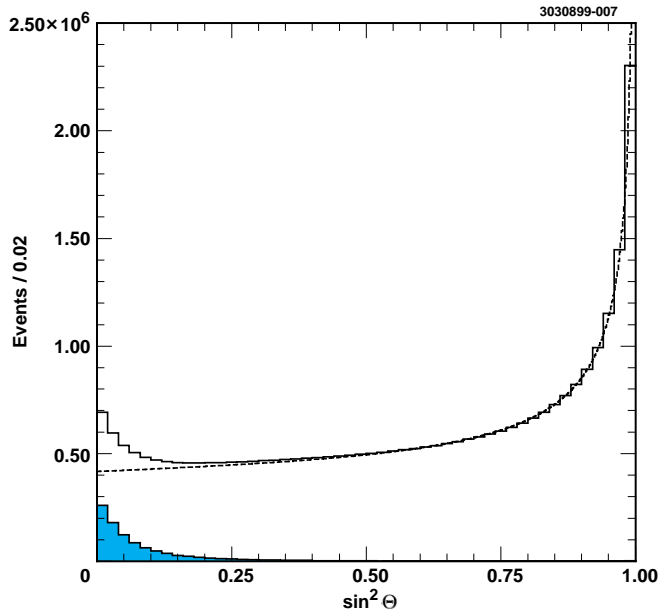}
\end{picture} \\
\caption{\label{fig:pion_sqrsine}
Shown is the inclusive 
${\rm sin}^2\theta$
distribution for all tracks (solid
histogram) overlaid with the background fit function (dashed) and the 
                $\pi^-_{\rm soft}$
signal expected from $D^{*-}\to{\overline D^0}\pi^-_{\rm soft}$ 
decays (shaded). 
Determination of signal and background follows
an earlier CLEO analysis[12],
which used this method to measure
${\cal B}(D^0\to K^-\pi^+)$.}
\end{figure}

(d) The pion's trajectory must lie near the trajectory of the parent
charm quark, as expected for pions produced in
$D^{*-} \to \overline{D}^0 \pi^-_{\rm soft}$.
Experimentally, this is checked using the variable 
${\rm sin}^2\theta$, where $\theta$ is the opening angle between 
the candidate soft pion
and the event thrust axis\cite{thrust}. 
Assuming that the thrust axis approximates
the original $c{\overline c}$ axis, true $\pi^-_{\rm soft}$ 
should populate the region
${\rm sin}^2 \theta \to 0$.  Fig.~\ref{fig:pion_sqrsine}
displays the soft pion ${\rm sin}^2 \theta$
distribution for candidates passing our event and track selection criteria.
The excess in the region ${\rm sin}^2 \theta \to 0$ 
constitutes our charm-tagged
sample.

\subsubsection{${\overline D^0}$, $D^-$, and $D_{\rm s}^-$ Tags}
Fully reconstructed
${\overline D}$-meson tags are detected in the modes
${\overline D^0}\to K^+\pi^-$,
$D^-\to K^+\pi^-\pi^-$, 
and $D_{\rm s}^-\to\phi\pi^-$. In all cases, final
state particles are required to pass 
DOCA criteria with respect to the primary vertex in both the
radial ($|{\rm DOCA}|<$5 mm) and beam ($|{\rm DOCA}|<$5 cm) coordinates. Final
state particles are also required to have 
specific ionization and time of flight
information
consistent with their assumed identities.

\subsection{Antiproton Tags}
To be considered as candidates for antiproton (or proton, in the
charge conjugate case) ``tags'',
charged particles detected in the central drift chamber must 
also pass
strict particle identification criteria. Using the available time-of-flight
and drift chamber specific
ionization measurements for each track, the likelihood
that a particle be an 
antiproton must be at least nine times larger than the
likelihood that the particle be a $K^-$ or a $\pi^-$. 
Antiproton tag candidates
must also pass the same vertex requirements as soft pion and electron
candidates. These vertex criteria help suppress backgrounds from
non-primary antiprotons (from ${\overline\Lambda}\to{\overline p}\pi^+$,
e.g.) or baryons generated by collisions of beam
particles with either
the beampipe itself or residual gas within the beampipe. 

It is important that our antiproton tags be direct, and not hyperon
daughters.
By combining
our antiproton candidates with remaining charged tracks in the same
event (assumed to be pions), we can reconstruct ${\overline \Lambda}$'s and
estimate the fraction of our antiproton tags which
are due to reconstructed ${\overline\Lambda}$'s
decay. We determine this fraction to be $<$2\% (Sect.
\ref{Xic-lambdabar}).

We check the fraction
of our proton tags originating in beam-gas
and beam-wall collisions by determining the asymmetry between the number
of proton tags and antiproton tags. If the beam-gas/beam-wall contamination
is large, we expect there to be a preponderance of proton tags compared to
antiproton tags. In fact, 
in a ${\overline D}$-meson tagged subset of the full data used in this
analysis,
we find the number of proton tags (6980$\pm$255)
to be 
statistically equal to the number of antiproton tags (6737$\pm$250).
Nevertheless,
the difference between these two numbers is 
taken as our systematic uncertainty in the magnitude of beam-related
backgrounds (Table II). 

\section{Triple Correlations}

In the triple correlation analysis,
we tag the ${\overline c}$ side of an $e^+e^-\to c{\overline c}$ event 
using a soft pion or an electron tag, then 
search for a $\overline{p}$ in the opposite hemisphere.  In order to conserve 
both charm and baryon number we assume a $\Lambda_c^+$ in the hemisphere 
opposite the tag.
Below we show a
schematic diagram of an
event where either a $\pi_{\rm soft}^-$ or $e^-$,
in combination with  
an anti-proton, is used to tag an unseen ($\Lambda_c^+$) decay.

\vspace{.25cm}

\begin{center}
\[\begin{array}{cccccccc}

& & & c &  \overline{c} & & & \\
& & \overline{p}\hookleftarrow & & &  \hookrightarrow 
D^{*-} & & \\ 
& & (\Lambda_c^+)\hookleftarrow & & & &  
\hookrightarrow \overline{D}^0 \pi_{\rm soft}^- & \\
&  (anything) \hookleftarrow & & & & & & \hookrightarrow e^- K^+ \nu_e \\ 

\end{array}\]
\end{center}

\vspace{.25cm}

\begin{figure}
\begin{picture}(200,250)
\includegraphics{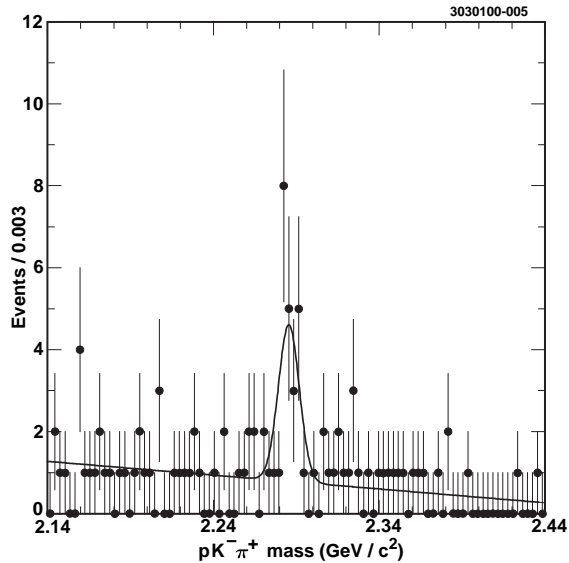}
\end{picture} \\
\caption{\label{fig:elamc}
         \small The 
candidate
$\Lambda_c^+$ mass (i.e., $pK^-\pi^+$ mass, in GeV/$c^2$) 
for $\Lambda_c^+$'s with a 
$\overline{p}$ in the same hemisphere [$\Lambda_c^+{\overline p}$]
and an $e^-$ in the opposite hemisphere ($\Lambda_c^+e$). The triple
correlation yield
is $10.3\pm3.8$ events.}
\end{figure}

The above diagram gives us a known sample of $\Lambda_c^+$ events.  
(Note that we do not require that both
$\pi_{\rm soft}^-$ and $e^-$ tags
be present in a candidate event; the presence of either one
constitutes a valid ``charm-tag''.)
In the electron tag case, the total number of $\Lambda_c^+$'s is the
number of events in which a track passes our electron tag identification and an
antiproton tag is found in the opposite hemisphere.
We then reconstruct, in that sample, a $\Lambda_c^+$ decaying 
into $p K^- \pi^+$ in the same hemisphere as the
${\overline p}$, and opposite the 
electron candidate. 
The $\Lambda_c^+$ invariant mass distribution
is then fit to a first order Chebyschev Polynomial 
to represent
the background and a Gaussian to represent
the signal (Fig. \ref{fig:elamc}), with the 
$\Lambda_c^+$ mass and width 
fixed to the values obtained from a fit to the 
inclusive $\Lambda_c^+$ mass spectrum in data.

When using the soft pion tag, we select events that are supposed
to contain a $\Lambda_c^+$ by plotting
the ${\rm sin}^2\theta$ distribution 
of pions with a 
tag $\overline{p}$ in the opposite hemisphere, with $\theta$ defined 
as before as the 
angle between the pion's momentum and the thrust axis 
(Fig. \ref{fig:pion_sqrsine}).  Background and 
signal distributions are then fit to this ${\rm sin}^2\theta$ 
distribution. The background function we use is 
$f(x) = C_1 (1/\sqrt{1-x}) + C_2 (1/\sqrt{1+Ax^2+Bx^3})$, 
where $x$ is ${\rm sin}^2 \theta$. This functional form is 
taken from a previous
CLEO measurement of ${\cal B}(D^0\to K^-\pi^+)$ using a
similar technique\cite{CLEO-D0toKpi}.

\begin{figure}
\begin{picture}(200,250)
\includegraphics{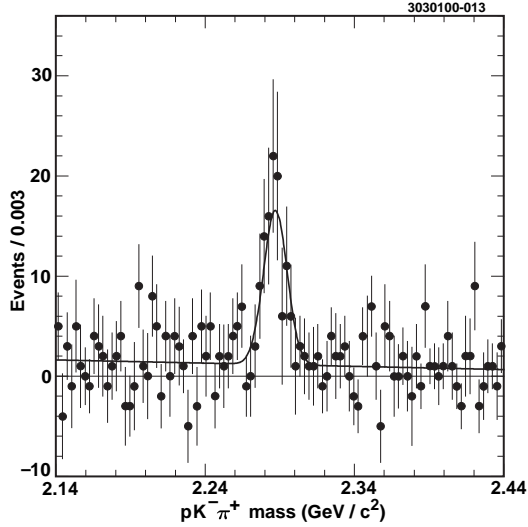}
\end{picture} \\
\caption{\label{fig:pionlamcfit}
         \small Results of sideband subtraction in data to determine 
$\Lambda_c^+\to pK^-\pi^+$ yield in soft-pion tagged events.
We project onto the
candidate $\Lambda_c^+$ mass axis the portion of our 
two-dimensional $pK^-\pi^+$ mass vs. ${\rm sin}^2\theta$
plot corresponding to 
${\rm sin}^2\theta < 0.25$ and subtract the 
scaled projection corresponding to $0.25\leq{\rm sin}^2\theta\leq 0.5$.  
We then perform a fit to the 
resulting $pK^-\pi^+$ mass spectrum 
in order to find our final yield of 
$c\overline{c} \to \Lambda_c^+ + \overline{p} + \pi_{\rm soft}^- + X$ 
events. The raw triple correlation yield is $101.6\pm20.6$ events.}
\end{figure}

Using the soft pion tag,
we extract the number of signal events from
a two-dimensional 
plot of $pK^-\pi^+$ invariant mass 
versus the ${\rm sin}^2\theta$ of the 
$\pi_{\rm soft}^-$  
From this two-dimensional 
distribution, we perform a 
scaled sideband
subtraction of the $\Lambda_c^+$ yield in the ``sideband'' region
($0.25<{\rm sin}^2\theta<0.5$) compared with the signal region 
(${\rm sin}^2\theta<0.25$) to determine the final,
background-subtracted yield
(Fig. \ref{fig:pionlamcfit}). (The background is approximately linear 
through this
region.) We have compared the yield obtained this way with the
yield obtained using
the ${\rm sin}^2\theta$ signal remaining after $\Lambda_c^+$-mass sideband
subtraction ($91\pm18$ events). The two techniques give consistent results;
the difference between them is counted towards the final systematic error
(Table II).

For both tags,
we can now quantify the 
ratio of tagged events containing a $\Lambda_c^+$ decaying into 
$pK^-\pi^+$ to all tagged 
events.  This ratio is equal to

\begin{center}
$$
R = \frac{{\cal Y}(e^+e^- \to \overline{D} + \overline{p} + 
(\Lambda_c^+ \to pK^-\pi^+) 
+ X)}{{\cal Y}(e^+e^- \to \overline{D} + \overline{p} + X)}
\eqno(1)$$
\end{center}

\noindent where ${\cal Y}$ stands for ``yield in $e^+e^-$ annihilation'',
``${\overline D}$'' designates any one of our charm tags, and the 
$\overline{p}$ and the $\overline{D}$ are in opposite hemispheres with 
respect to the thrust axis of the event.

Now the numerator can be written as

\begin{center}
$$
{\cal Y}(e^+e^- \to \overline{D} + \overline{p} + (\Lambda_c^+ \to pK^-\pi^+) 
+ X) =
$$
$$
{\cal L} \cdot \sigma(e^+e^- \to c\overline{c}) \cdot 
{\cal B}(c\overline{c} \to \overline{D} + 
\overline{p} + \Lambda_c^+ + X) \cdot {\cal B}(\Lambda_c^+ \to pK^-\pi^+) \cdot 
(\epsilon_{\overline{D}}) \cdot (\epsilon_{\overline{p}}) \cdot 
(\epsilon_{\Lambda_c^+})
\eqno(2)$$
\end{center}

\noindent{and the denominator}

\begin{center}
$$
{\cal Y}(e^+e^- \to \overline{D} + \overline{p} + X) = {\cal L} \cdot 
\sigma(e^+e^- \to c\overline{c}) \cdot {\cal B}(c\overline{c} \to 
\overline{D} + \overline{p} + \Theta_c + X^{\prime}) \cdot 
(\epsilon_{\overline{D}}) \cdot (\epsilon_{\overline{p}})
\eqno(3)$$
\end{center}

\noindent where $\Theta_c$ is any charm+baryon system, not necessarily 
a $\Lambda_c^+$
(e.g. it could be a $D$ + nucleon or a charmed baryon, such as a 
$\Xi_c$, not always decaying into $\Lambda_c^+ + X$),
${\cal L}$ is the total luminosity, and $\epsilon_{\overline p}$,
$\epsilon_{\Lambda_c^+}$, and 
$\epsilon_{\overline{D}}$ are the efficiencies of 
finding the antiproton, $\Lambda_c^+$, and charm tags, respectively.

We then write

\begin{center}
$$
{\cal Y}(e^+e^- \to \overline{D} + \overline{p} + X) = f_1 \cdot 
{\cal L} \cdot \sigma(e^+e^- \to c\overline{c}) 
\cdot {\cal B}(c\overline{c} \to \overline{D} + \overline{p} + 
\Lambda_c^+ + X) \cdot \epsilon_{\overline{D}} \cdot \epsilon_{\overline{p}}
\eqno(4)$$
\end{center}

\noindent where

\begin{center} 
$$
f_1 \equiv \frac{{\cal B}(c\overline{c} \to \overline{D} + \overline{p} + 
\Theta_c + X^{\prime})}{{\cal B}(c\overline{c} \to \overline{D} 
+ \overline{p} + \Lambda_c^+ + X)}.
\eqno(5)$$
\end{center}

\noindent Since $f_1$ takes into account the fact that our
yield includes also charmed, baryonic systems other than $\Lambda_c$,
$f_1\geq$1.0. Then:

\begin{center}
$$
{\cal B}(\Lambda_c^+ \to pK^-\pi^+) = 
\frac{R \cdot f_1}{\epsilon_{\Lambda_c^+}}.
\eqno(6)$$
\end{center}

Since the above equation holds for both data and Monte Carlo simulations
we can write:

\begin{center}
$$
\frac{{\cal B}(\Lambda_c^+ \to pK^-\pi^+)_{Data}}{{\cal B}(\Lambda_c^+ \to 
pK^-\pi^+)_{MC}} = \frac{R_{Data} \cdot \epsilon_{\Lambda_c^+ (MC)}
\cdot f_{1(Data)}}{R_{MC} \cdot \epsilon_{\Lambda_c^+ (Data)} 
\cdot f_{1(MC)}}.
\eqno(7)$$
\end{center}

We use Monte Carlo simulations to determine event and particle
reconstruction efficiencies. 
The simulated sample size corresponds to approximately 6 ${\rm fb}^{-1}$
of integrated luminosity.
Our Monte Carlo simulation combines an 
$e^+e^-\to c{\overline c}$
event generator
(JETSET 7.3\cite{JETSET}) with a GEANT-based\cite{GEANT} 
simulation of our detector.
Assuming that the detector simulation
accurately reproduces the efficiency of reconstructing a 
$\Lambda_c^+$ in a tagged event and that 
we can determine the correction $f_1$ in both data and Monte Carlo,
we can then 
calibrate our observed value of
$\Lambda_c^+$ per tagged event in data to Monte Carlo:

\begin{center}
$$
\frac{f_{1,Data}}{f_{1,MC}}\frac{R_{Data}}{R_{MC}} 
\cdot {\cal B}(\Lambda_c^+ \rightarrow pK^-\pi^+)_{MC} =
{\cal B}(\Lambda_c^+ \rightarrow pK^-\pi^+)_{Data}.
\eqno(8)
$$
\end{center}


\vspace{1cm}
\subsection{Purity of Our Event Sample{\label{purity}}}

\begin{figure}
\begin{picture}(200,250)
\includegraphics{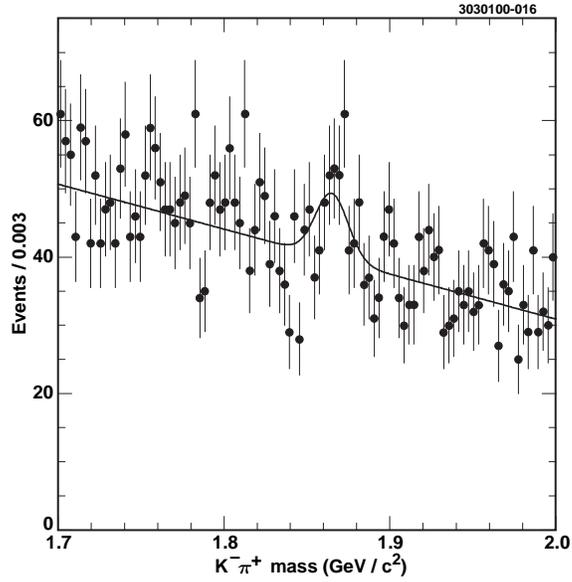}
\end{picture} \\
\begin{picture}(200,250)
\includegraphics{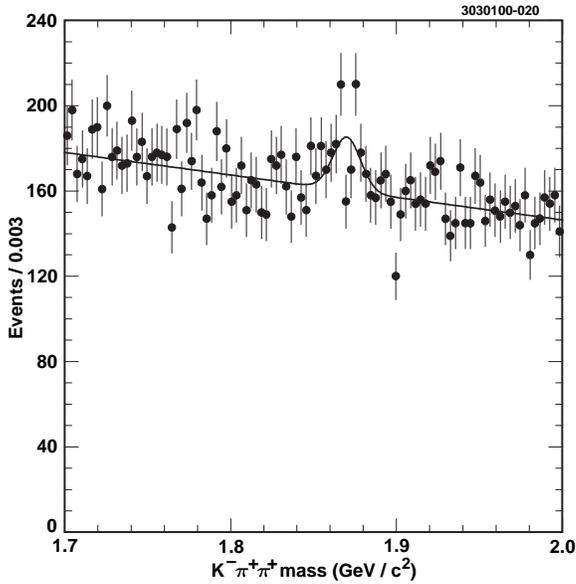}
\end{picture} \\
\caption{\label{fig:pd0}
         \small Candidate
$D^0$ (top) and $D^+$ (bottom) mass (GeV/$c^2$) 
for $D$ candidates in the same hemisphere as a $\overline{p}$ 
[$D{\overline p}$].  Events in the $D$ signal region 
are $D\overline{D}N\overline{p}$ events that contaminate our 
candidate
$\Lambda_c^+$ event sample.  The masses and widths of the
$D^0$ and $D^+$ are taken from fits to the  
inclusive mass spectra in data.} 
\end{figure}

We seek, wherever possible, to measure backgrounds directly from data and
thereby minimize the Monte Carlo dependence; i.e., we prefer to
measure $f_{1(Data)}$ and $f_{1(MC)}$ separately rather than to assume
equality of these fractions.
According to event simulations, the primary 
non-$\Lambda_c^+$ contribution to the numerator of $f_1$
is due to events where baryon number opposite the
${\overline p}$ tag is conserved by another
nucleon and a $D$ meson is created in the hemisphere opposite our anticharm
(``${\overline D}$'') tag, so that no $\Lambda_c^+$ is present in the event.
We refer to these events as
$D{\overline D}N{\overline p}$ events.
In order
to estimate the number of $D\overline{D}N\overline{p}$ events that 
contaminate our 
tagged event sample, we measure the number of events containing a 
$\overline{p}$ tag and a $D$ meson in the same hemisphere 
(S[${\overline p}D$], see Figure 
\ref{fig:pd0}) and 
assume (without reconstructing) 
a $\overline{D}$ in the opposite hemisphere to conserve
charm.  A 
correction (11$\pm$2\% in data, described in Sections (\ref{purity})
and (\ref{purity2}) of this
document)
is made to ${\cal B}(\Lambda_c^+ \to pK^-\pi^+)$ based on the 
observed yields of these $D\overline{D}N\overline{p}$ events
in data and Monte Carlo.  
A much smaller contribution
to $f_1$ arises from $\Xi_cX{\overline D}{\overline p}$ 
(also discussed later in the text).

Anti-protons 
from $\overline{\Lambda}_c$ decay 
entering the hemisphere of the $D$ meson and 
$\overline{\Lambda}_c\overline{p}D$ events must not be large in order 
for our assumption that the S$[D{\overline p}]$ sample can be used to
estimate the level of $D{\overline D}N{\overline p}$ background be valid.
In order to check for anti-protons from 
$\overline{\Lambda}_c$'s decaying into the hemisphere of a $D$ we plot the 
cosine of the angle between the anti-proton's momentum vector and its parent 
$\overline{\Lambda}_c$. For anti-protons passing our event and
track criteria, back hemisphere leakage is found to be
negligible ($<1\%$).  Events containing 
$\overline{\Lambda}_c\overline{p}D$ must contain two baryon-anti-baryon 
pairs as well as a charmed meson 
(e.g. ${\overline\Lambda_c^-}\overline{p}NND$).  
Although it is possible to have four baryons and a charmed meson in 
the same event it should be noted that this process would lead to an 
overestimation of our background 
(i.e. events that contain a $D \overline{p}$ but do not tag 
$D\overline{D}N\overline{p}$ events), thus biasing us towards 
a ${\cal B}(\Lambda_c^+ \to pK^-\pi^+)$ that is higher than the 
true branching fraction. Monte Carlo simulations indicate that this
background is exceedingly ($<$1\%) small.

\subsubsection{Contamination of the $\pi^-_{\rm soft}$ sample}
\begin{figure}
\begin{picture}(200,250)
\includegraphics{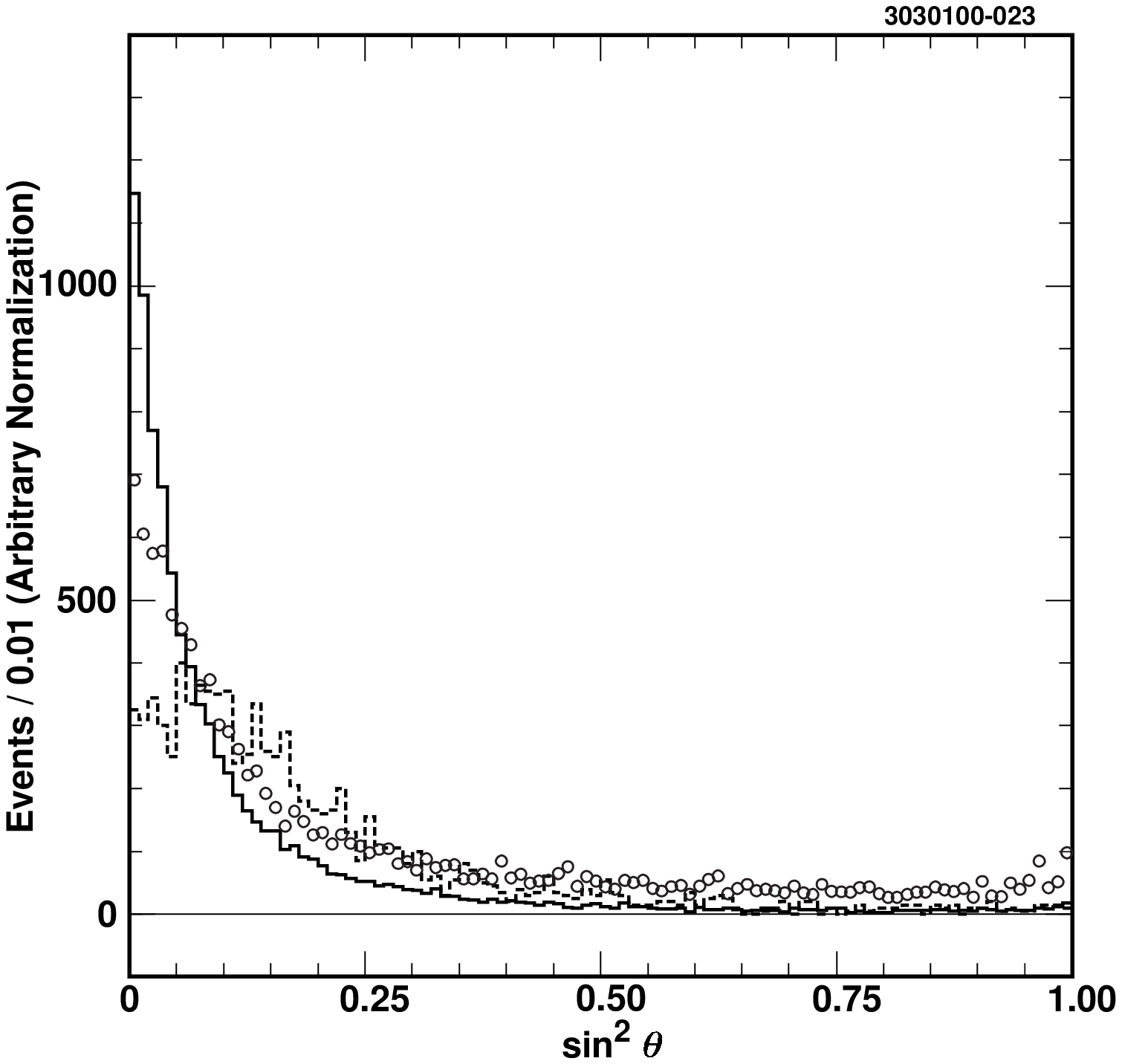}
\end{picture} \\
\caption{\label{fig:sin2overlay}
         \small Monte Carlo ${\rm sin}^2\theta$ distribution of $\pi^-$'s from 
$D^{*-}\to D^0\pi^-$ (solid line), $\Lambda_{cJ}^+(2593) \to 
\Lambda_c^+\pi^+\pi^-$ (diamonds), and $\Sigma_c^0 \to \Lambda_c^+\pi^-$ 
(dashed line) after all event and particle identification cuts 
are applied to the $\pi^-$'s.}
\end{figure}

\begin{figure}
\begin{picture}(180,220)
\includegraphics{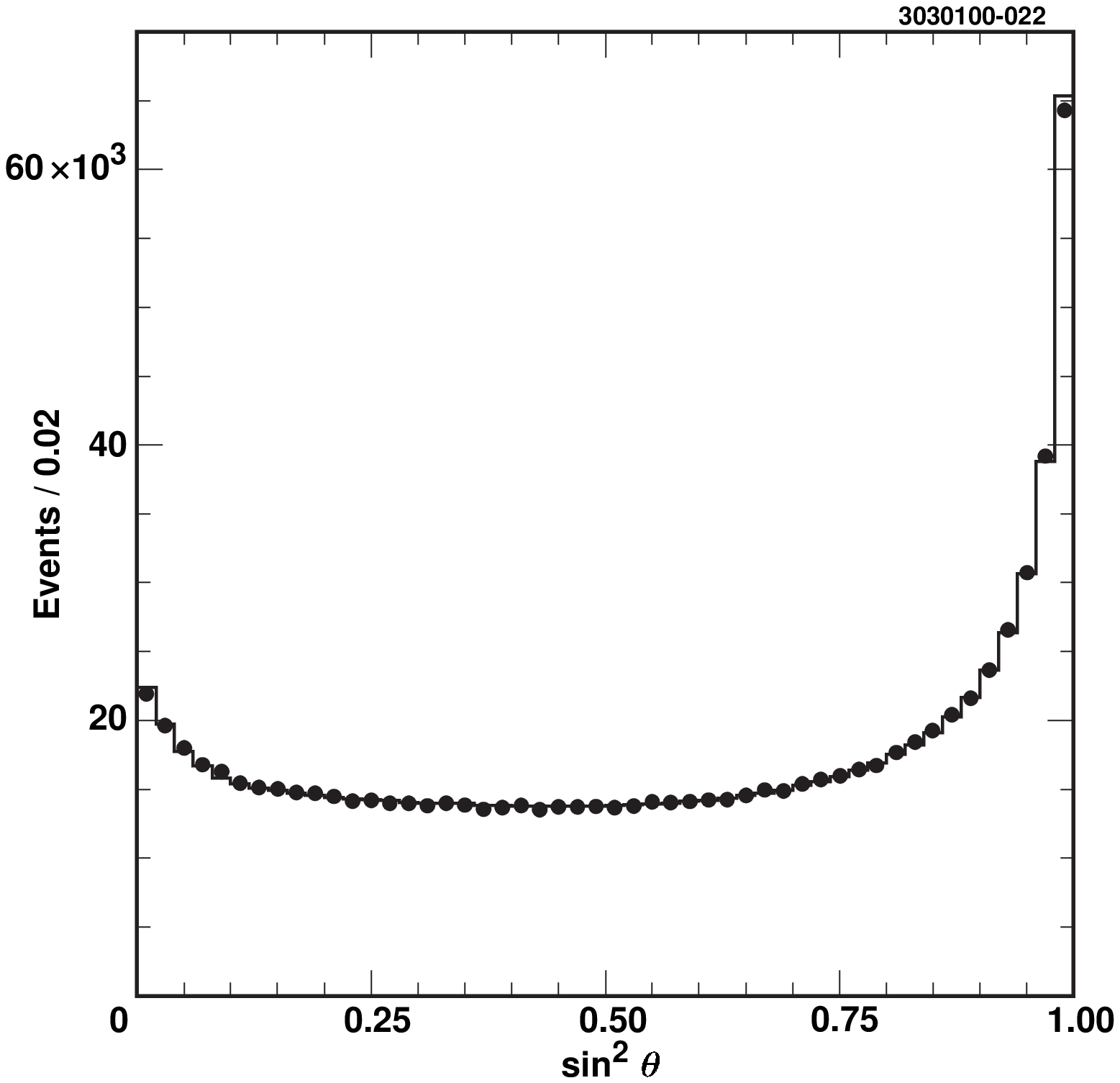}
\end{picture} 
\begin{picture}(180,220)
\includegraphics{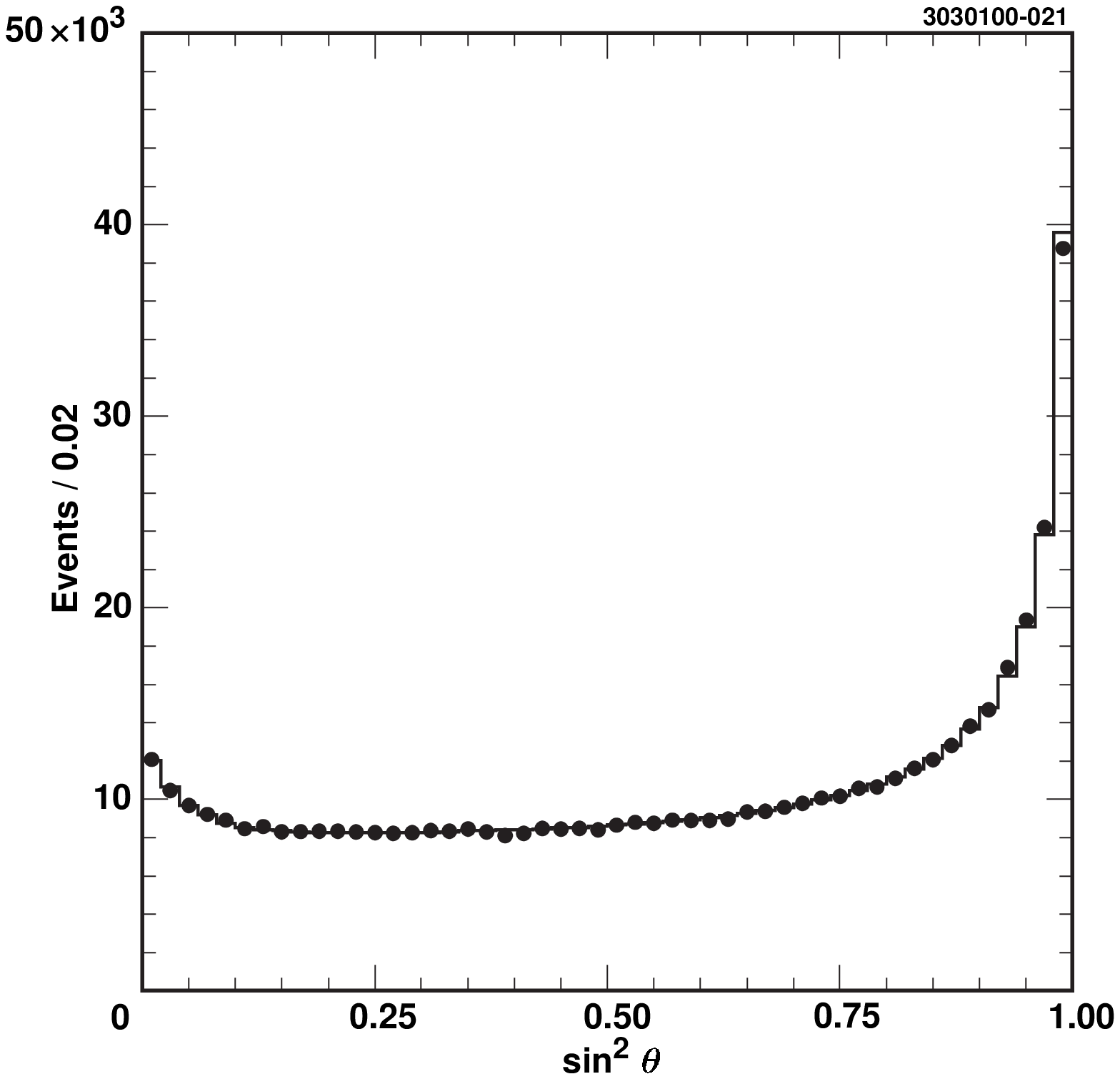}
\end{picture} \\
\caption{\label{fig:sigmacdata}
         \small ${\rm sin}^2\theta$ distribution of $\pi^-$'s in
data events containing 
a tag antiproton in the opposite hemisphere 
O($\pi^-_{\rm soft}|{\overline p}$), for
Monte Carlo simulations (left) vs. data (right).  
A free fit is performed using 
the Monte Carlo ${\rm sin}^2\theta$ distributions for $\pi^-$'s decaying from 
$D^{*-}\to D^0\pi^-$, $\Sigma_c^0\to\Lambda_c^+\pi^-$, 
and $\Lambda_c(2593)\to\Lambda_c^+\pi^+\pi^-$. 
This plot is made after all event and
$\pi_{\rm soft}^-$ particle
identification cuts have been applied. The fitted
$\Sigma_c+\Lambda_{cJ}$ fractions for Monte Carlo and data
are $21\pm9$\% and $14\pm17$\%, respectively.} 
\end{figure}

Pions from $\Sigma_c^0 \to \Lambda_c^+\pi^-$ and orbitally
excited $\Lambda_{cJ}^+ \to 
\Lambda_c^+\pi^+\pi^-$ decays have ${\rm sin}^2\theta$ 
distributions similar to the soft pions from $D^{*-}$ decays as seen in 
Figure \ref{fig:sin2overlay}.  
Although the number of 
$\Sigma_c^0$ and $\Lambda_{cJ}^-$ particles (primarily 
$\Lambda_c^-(2593)$ and $\Lambda_c^-(2630)$) is small 
relative to the number of 
$D^{*-}$ particles, this background is
potentially significant since the likelihood for 
having a $\overline{p}$ tag is large in events containing these charmed 
baryons.  In order to estimate the magnitude of these events in data and 
Monte Carlo we perform a fit using 
Monte Carlo-derived ${\rm sin}^2\theta$ distributions for 
tagged
$\pi^-_{\rm soft}$ decaying from both $D^{-*}$ and $\Sigma_c^0$ decays.  
We fit these 
distributions to our plot of
the inclusive $\pi^-_{\rm soft}$ ${\rm sin}^2\theta$ spectrum in events 
containing a $\overline{p}$ in the
hemisphere opposite the $\pi^-_{\rm soft}$ 
with respect to the thrust axis (see Figure 
\ref{fig:sigmacdata}).  The difference  
between the data and Monte Carlo ($\Sigma_c + 
\Lambda_{c,J}$) $\pi^-_{\rm soft}$ fit fractions 
relative to the total 
$\pi^-_{\rm soft}$ yield
in data (14 $\pm$ 17)\% as compared to 
Monte Carlo (21 $\pm$ 9)\%\footnote{The actual
fraction in Monte Carlo is 12\%.}
is taken as a systematic error (Table II).

\subsubsection{Electron Tag Backgrounds}
We assume that our tag electrons are not only true electrons in 
$c\overline{c}$ events, but also that they are coming from semileptonic
charm decay.
In Monte Carlo, 
$\sim$ 87\% of our tag electrons are true electrons coming from 
charm semileptonic decays.
The remainder of
our tag electrons are either background fakes (i.e.,
non-electrons) or background electrons not from charm
decays (predominantly from the decay $\pi^0 \to e^+e^-\gamma$).
Each of these backgrounds contributes approximately equally to
our candidate electron sample. 
The number of fake electron tags should cancel in our 
equation for ${\cal B}(\Lambda_c^+ \to pK^-\pi^+)$, unless there is a 
decreased probability of tag electron fakes in events that contain a 
$\Lambda_c^+\overline{p}$ as compared to those only containing a 
$\overline{p}$.  Since this very well may be the case, we vary the 
electron identification cuts and take the change in the calculated 
${\cal B}(\Lambda_c^+ \to pK^-\pi^+)$ as a systematic error (5\%,
as listed in Table II).

Another possible source of tag electron background is from two-photon 
annihilations, in which one of the incident beam particles
scatters into the detector.  The two-photon contamination is assessed by 
determining the asymmetry between the number of positrons in the 
forward hemisphere compared to the number of electrons in the negative
hemisphere (beam positrons define $+{\hat z}$ in the local coordinate
system).
We find two-photon annihilations to be negligible ($<1\%$) in our tag 
electron sample.

\subsubsection{Backgrounds from $([\Xi_c{\overline\Lambda}]|{\overline D})$}
\label{Xic-lambdabar}
\begin{figure}
\begin{picture}(200,250)
\includegraphics{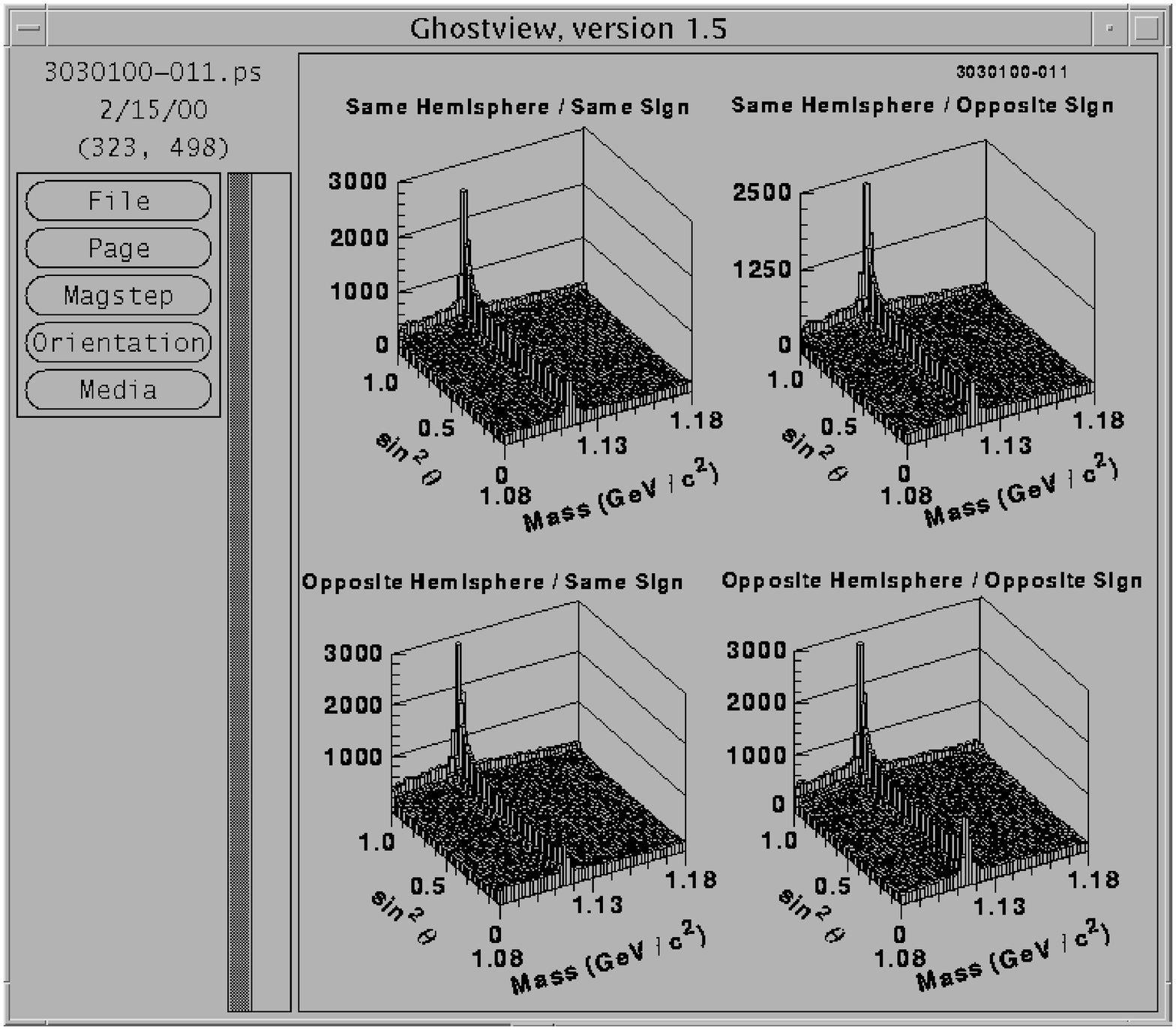}
\end{picture} \\
\caption{\label{fig:lamcorr}
         \small 
Candidate $\overline{\Lambda}$ mass versus the ${\rm sin}^2\theta$ 
of tag $\pi_{\rm soft}^-$ in the opposite hemisphere in
data O(${\overline\Lambda}|\pi^-_{\rm soft}$).  The lower 
left hand plot is the hemisphere and sign correlation 
(opposite hemisphere/same sign) of interest.  We use this
plot to check
for an excess of $([\Xi_c\overline{\Lambda}]|D^{*-})$ events in data 
as compared to Monte Carlo. The excess of candidate signal events at
${\rm sin}^2\theta\to$0 and 
$m_{{\overline p}\pi^-}\sim m_{\Lambda}$ in the lower
right-hand plot is attributed to O($\Lambda_c^+|D^{*-}$) events, in which
$\Lambda_c^+\to\Lambda X$.}
\end{figure}

Tagged events may also contain a charmed baryon other than a $\Lambda_c^+$; 
most likely a $\Xi_c$.  It is therefore important 
to check that the ratio of $\Xi_c/\Lambda_c^+$ production
rates is similar in data 
and Monte Carlo simulations. 
Monte Carlo simulations (JETSET 7.3) indicate that, in events passing
our event selection criteria, and having an
antiproton tag originating from the primary vertex,
$\Xi_c/\Lambda_c^+$=0.014.  Since this fraction is so small in 
Monte Carlo simulations,
the data fraction must be inconsistent with the Monte Carlo 
expectation
by at 
least an order of magnitude to make a significant difference in our 
calculation of ${\cal B}(\Lambda_c^+ \to pK^-\pi^+)$.  In order to 
check the fraction of our tagged event sample containing a $\Xi_c$ 
instead of a $\Lambda_c^+$, we plot 
the ${\rm sin}^2\theta$ of $\pi_{\rm soft}^-$ 
versus the mass of an opposite hemisphere 
$\overline{\Lambda}$ (Figure \ref{fig:lamcorr}), rather
than an opposite hemisphere ${\overline p}$.  This
is not the 
correct sign correlation for $\overline{\Xi}_c\to\overline{\Lambda}$ 
decays since the $\pi_{\rm soft}^-$ tags a $D^{*-}$ (such
a correlation implies ${\overline
c}$ in both hemispheres).  Instead, we assume 
that the 
dominant contributor to this plot is from $\overline{\Lambda}$'s 
conserving baryon number with a same hemisphere $\Xi_c$ 
(i.e. $([\Xi_c\overline{\Lambda}]|D^{*-})X$ events).  Although 
other O(${\overline\Lambda}|{\overline D}$) topologies may contribute
(e.g. ${\overline D^0}\overline{\Lambda}K^-N$), it is still probable 
that an excess of $\Xi_c$ production in our $\pi^-_{\rm soft}$ event sample 
would be noticed as an excess in $\overline{\Lambda}$ production opposite 
our tag $\pi_{\rm soft}^-$.  In fact we do not see this excess; we find
$(1.3\pm0.2)\times 10^{-3}$ ${\overline\Lambda}$ per 
$\pi_{\rm soft}^-$ tagged event in data vs. 
$(1.6\pm0.2)\times 10^{-3}$ in Monte Carlo simulations.
(Note that we 
have already suppressed $\Xi_c\overline{\Lambda}D^{*-}$ events by 
requiring the tag anti-proton to come from the 
primary event vertex.)  These checks do 
not however address a possible excess of $\Xi_c{\overline p}KD^{*-}$ events.  
In principle one could estimate $\Xi_c/\Lambda_c$ in data by 
reconstructing $\Xi_c$'s in the hemisphere of a tag anti-proton.  However, 
we are unable to make an accurate estimate of the $\Xi_c{\overline p}$ 
yield due to low reconstruction efficiency.
Therefore, using
the similarity of the relative
${\overline\Lambda}/\pi^-_{\rm soft}$ production ratio in
data vs. Monte Carlo simulations as guidance,
we assume the same $\Xi_c/\Lambda_c$ production ratio in data
as in Monte Carlo simulations. 
It should be noted that 
although the systematic error assessed 
due to uncertainties in $\Xi_c$ production is 
small (3\%, see Table II), this magnitude of 
systematic error represents twice the amount of $\Xi_c$ production 
predicted by Monte Carlo simulations for
our tagged event sample.

\subsection{Hemisphere Correlation}
\begin{figure}
\begin{picture}(200,250)
\includegraphics{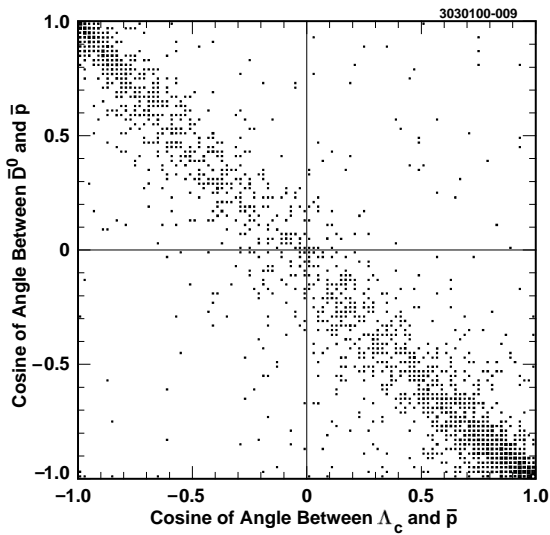}
\end{picture} \\
\caption{\label{fig:hemispheres}
         \small Cosine of the angle between the tag anti-proton momentum 
vector
and $\Lambda_c^+$ momentum vector (horizontal) 
versus cosine of the angle between the tag 
anti-proton momentum and $\overline{D}^0$ momentum (vertical) in 
$\Lambda_c^+\overline{D}^0\overline{p}$ events from Monte Carlo
simulations, with no particle cuts (i.e., minimum momentum and
track reconstruction cuts, etc.)
applied. Events in the lower-left hand quadrant 
are due primarily
to $c\overline{c}g$ and $c\overline{c}\gamma$ events.  In 
these events we have an antiproton that passes our tag anti-proton cuts but 
is in the opposite hemisphere of both a $\Lambda_c^+$ and a $\overline{D}^0$, 
thus giving us a slight 
excess ($\sim$2\%) of $\overline{D}^0$ opposite $\overline{p}$ events 
relative to $\Lambda_c^+\overline{p}$ same hemisphere events.  
The lower-right hand quadrant 
corresponds to our signal events. The upper-left hand quadrant event sample
is used later for a cross-check (Sect VD).}
\end{figure}

There is an additional systematic
error due to the hemisphere correlation
requirements we impose on the (${\overline p}|{\overline D}$) 
and [${\overline p}\Lambda_c^+$] samples.
In fact, not all ${\overline p}{\overline D}\Lambda_c^+$ 
events in which the ${\overline p}{\overline D}$ are in
opposite hemispheres 
necessarily have the ${\overline p}\Lambda_c^+$ in the same
hemisphere (e.g., if the three momentum vectors have opening angles of
$120^\circ$ between them). This can happen in events with photon or gluon
radiation. 
For $c{\overline c}g$ or $c{\overline c}\gamma$ 
(initial state radiation) 
events, the ${c\overline c}$ 
will not be directly back to back. This can be seen
in Figure \ref{fig:hemispheres}, which shows the 
${\overline p}{\overline D}$ opening angle
vs. the ${\overline p}\Lambda_c^+$ opening angle for Monte Carlo
${\overline p}{\overline D}\Lambda_c^+$ events. 
Note that, in producing this
distribution, we have not required that either the ${\overline D}$ or
$\Lambda_c^+$ be high momentum, as we would for our standard data analysis,
whereas the momenta for charm particles in radiative $c{\overline c}$
events is typically smaller. Nevertheless,
the fraction of $\Lambda_c^+{\overline p}{\overline D}$ events
in which the antiproton is found in the lower-left quadrant of
Figure \ref{fig:hemispheres} is taken as a
systematic error (Table II), reflecting the fact that
the hemisphere correlation is not rigorous, and that these 
angular distributions
may be different in data vs. Monte Carlo.

\section{Double Correlations}

\subsection{Method}

In order to circumvent the low statistics involved with the triple correlation 
methods we exploit a double correlation method.  We begin with events 
containing a S$[\overline{p}\Lambda_c^+]$ in the same hemisphere.  
In these 
events a $\overline{c}$-hadron can be assumed in the hemisphere
opposite 
the $\Lambda_c^+$.  This $\overline{c}$-hadron will most likely be an 
anti-charmed meson.  Events containing  
an anti-charmed baryon opposite a  $\overline{p}\Lambda_c^+$ should be 
suppressed due to the energy required to create four baryons in an event, as 
well as the small $c \to \Lambda_c$ fragmentation rate.  
Below is a representation of a sample 
event in which a $\overline{p}$ and a $\Lambda_c^+ \to pK^-\pi^+$ are observed 
with a $\overline{D}^0$ or $D^-$ or $D_{\rm s}^-$ 
assumed to exist in the opposite 
hemisphere.  

\vspace{.5cm}

\[\begin{array}{ccccccccc}
& & &  & c &  \overline{c}  & &  \\

& & &  \overline{p}\hookleftarrow & & & \hookrightarrow 
({\overline D}^0 \hspace{.25cm} or \hspace{.25cm} D^- 
\hspace{.25cm} or \hspace{.25cm} D_{\rm s}^-)  &   \\ 

& & & \Lambda_{c}^+\hookleftarrow & & & &  \hookrightarrow anything \\

& &  p  K^- \pi^+ \hookleftarrow & & & &   \\

\end{array}\]

\vspace{.5cm}

\begin{figure}
\begin{picture}(200,250)
\includegraphics{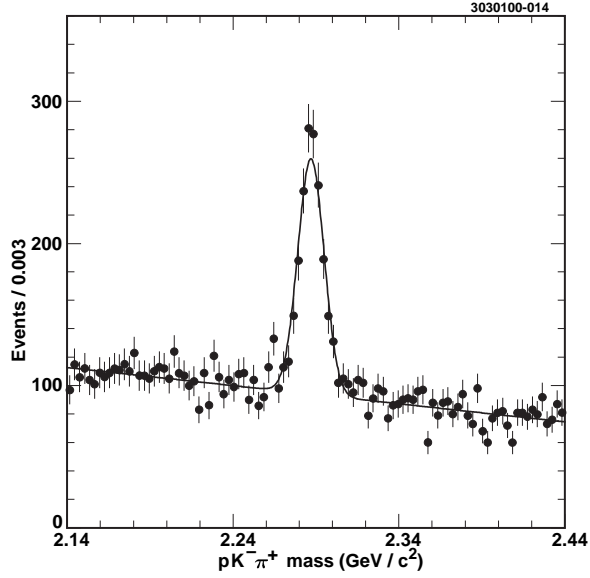}
\end{picture} \\
\caption{\label{fig:plamc}
         \small Candidate
$\Lambda_c^+$ mass (i.e., $pK^-\pi^+$ mass,
in GeV/$c^2$) for $\Lambda_c^+$'s with a 
$\overline{p}$ in the same hemisphere S[${\overline p}\Lambda_c^+$].  
In these events a $\overline{D}$ 
meson is assumed to recoil in the hemisphere opposite the 
$\Lambda_c^+$: O(${\overline D}|\Lambda_c^+$).}
\end{figure}

\begin{figure}
\begin{picture}(200,250)
\includegraphics{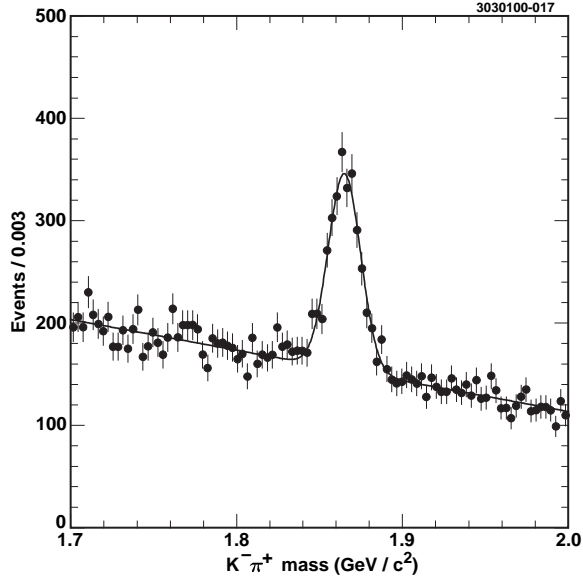}
\end{picture} \\
\caption{\label{fig:pd0bar}
         \small 
Candidate $\overline{D}^0$ mass (i.e., 
$K^+\pi^-$ mass, in GeV/$c^2$) with a $\overline{p}$ 
in the opposite hemisphere O(${\overline D}|{\overline p}$).  
In these events a $\Lambda_c^+$ is assumed 
to exist in the hemisphere opposite the $\overline{D}^0$: 
O(${\overline D^0}|\Lambda_c^+$).}
\end{figure}

\begin{figure}
\begin{picture}(200,250)
\includegraphics{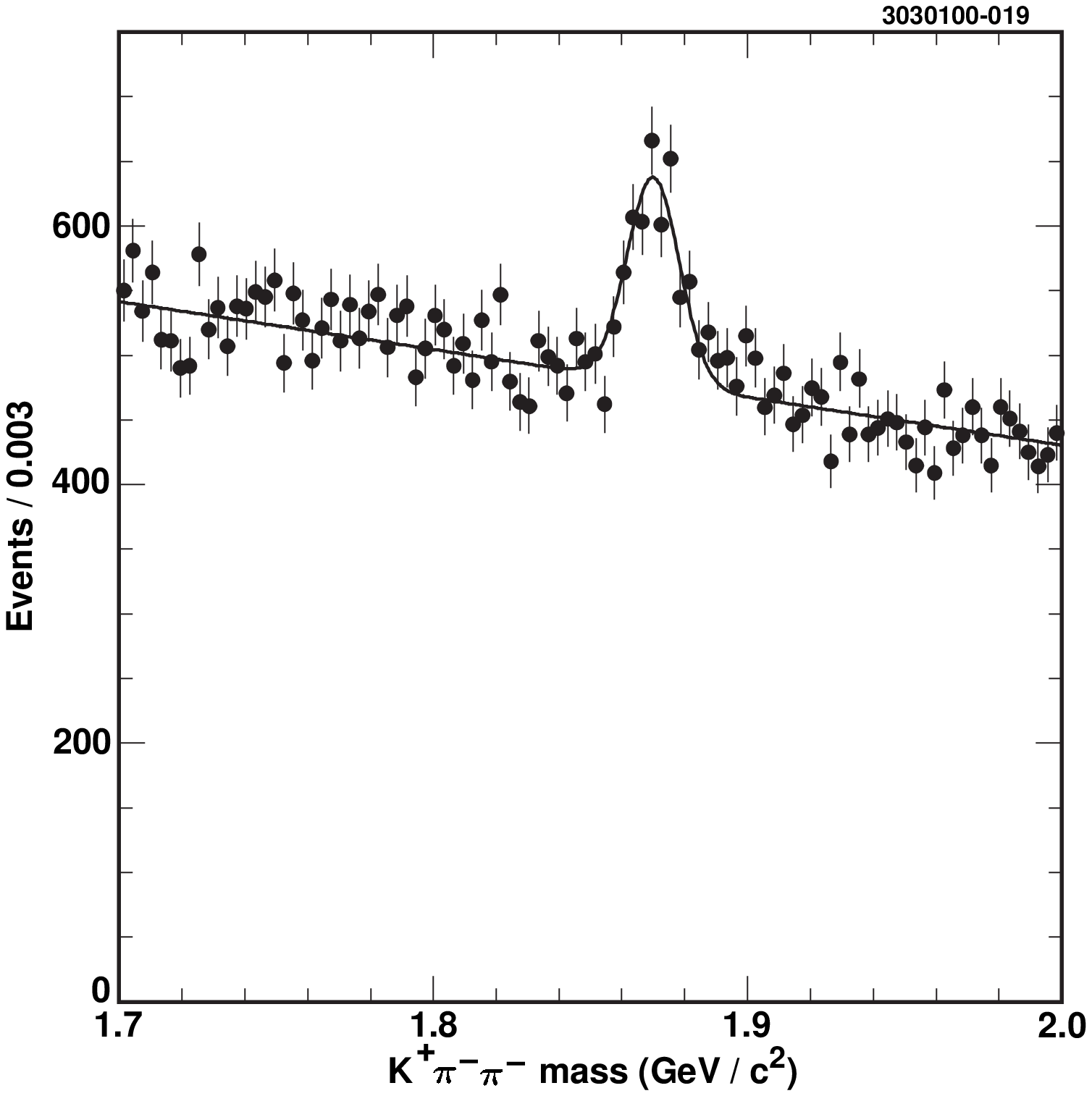}
\end{picture} \\
\caption{\label{fig:pDminus}
         \small Candidate
$D^-$ mass (i.e., $K^+\pi^-\pi^-$ mass, in GeV/$c^2$) with a $\overline{p}$ 
in the opposite hemisphere.  In these events a $\Lambda_c^+$ is assumed 
to exist in the hemisphere opposite the $D^-$: O($D^-|\Lambda_c^+$).}
\end{figure}

\begin{figure}
\begin{picture}(200,250)
\includegraphics{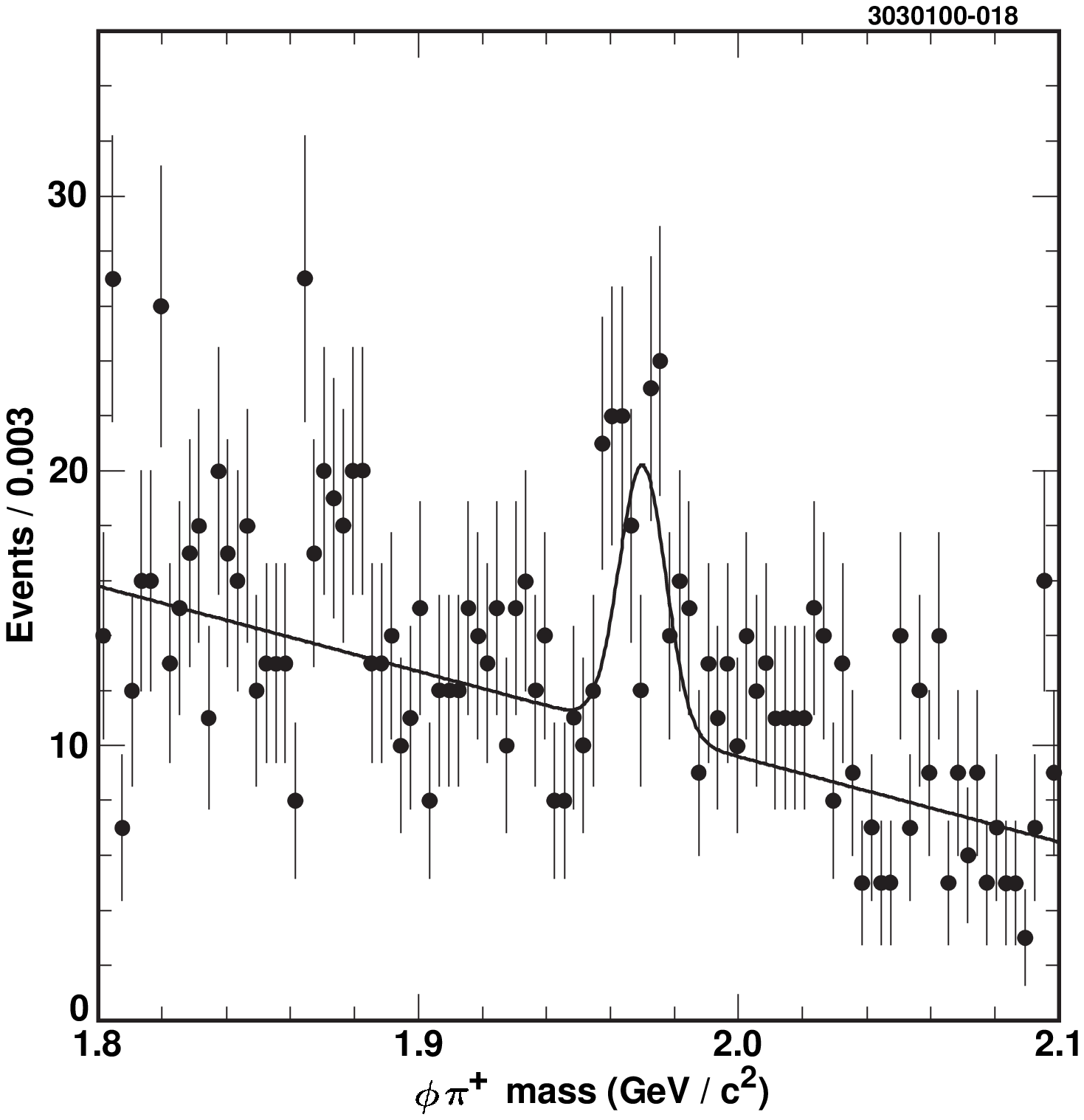}
\end{picture} \\
\caption{\label{fig:pDsminus}
         \small Candidate
$D_{\rm s}^-$ mass 
(i.e., $\phi\pi^-$ mass, in GeV/$c^2$) with a $\overline{p}$ 
in the opposite hemisphere.  In these events a $\Lambda_c^+$ is assumed 
to exist in the hemisphere opposite the $D_{\rm s}^-$:
O($D_s^-|\Lambda_c^+$).}
\end{figure}

After finding the number of 
S[$\overline{p}\Lambda_c^+$] events (Figure 
\ref{fig:plamc}), 
we separately find the 
number of times that a $\overline{p}$ is found opposite a $\overline{D}$.  
For this double correlation measurement, 
fully reconstructed mesons are used as
the anticharm tag $\overline{D}$.
We reconstruct the $\overline{D}^0$ (Figure \ref{fig:pd0bar}) through 
the $K^+\pi^-$ decay mode, the $D^-$ (Figure \ref{fig:pDminus}) through the 
$K^+\pi^-\pi^-$ decay mode, and the $D_{\rm s}^-$ (Figure \ref{fig:pDsminus}) 
through the $\phi \pi^-$ decay mode.  
We require the
${\overline D}$-meson 
to have momentum $p>$2.5~GeV/$c$, beyond the maximum possible in
$B\to {\overline D}$X events.
In these events we assume a $\Lambda_c^+$ in the hemisphere opposite
the $\overline{D}$.

\vspace{.5cm} 

\[\begin{array}{ccccccccc}
& & & & & c &  \overline{c}  & & \\

& & & & \overline{p}\hookleftarrow & & & \hookrightarrow 
\overline{D}^0 \hspace{.25cm} or \hspace{.25cm} D^- \hspace{.25cm} or 
\hspace{.25cm} D_{\rm s}^- & \\

& & & & (\Lambda_{c}^+)\hookleftarrow & & & &  
\hookrightarrow K^+ \pi^- \hspace{.25cm} or \hspace{.25cm} K^+ \pi^- \pi^- 
\hspace{.25cm} or \hspace{.25cm} \phi \pi^- \\

& & & anything \hookleftarrow \\
\end{array}\]

\vspace{.5cm}

Comparing the number of S[$\overline{p}\Lambda_c^+$] events to the number of 
O($\overline{p}|\overline{D}$) events, we are able to calculate 
$B({\Lambda_c^+ \to pK^-\pi^+})$, as follows.  First we write an equation for 
${\cal Y}(\Lambda_c^+)$, the yield of $\Lambda_c^+ \to pK^-\pi^+$ 
events containing a tag antiproton in the same hemisphere of the 
$\Lambda_c^+$:

\vspace{.5cm}

\begin{center}
$$
{\cal Y}[\Lambda_c^+\overline{p}] = 
\frac{{\cal L} \cdot \sigma(e^+e^- \to c\overline{c}) 
\cdot {\cal B}(c\overline{c} 
\to \overline{p} + \Lambda_c^+ + X + {\overline D}^0 
\hspace{.1cm} or \hspace{.25cm} D^- \hspace{.25cm} or 
\hspace{.1cm} D_{s}^-) \cdot {\cal B}(\Lambda_c^+ \to pK^-\pi^+) 
\cdot \epsilon_{\overline{p}} 
\cdot \epsilon_{[\Lambda_c^+\overline{p}]}}{1-f_2},
\eqno(9)$$
\end{center}

\noindent where $\epsilon_{[\Lambda_c^+\overline{p}]}$ is the efficiency for 
reconstructing a $\Lambda_c^+ \to pK^-\pi^+$ decay in an event containing a 
tag anti-proton and $f_2$ is defined as the fraction 
of $\Lambda_c^+\overline{p}$ events not containing a 
${\overline D}^0$ or $D^-$ or $D_{\rm s}^-$:

\begin{center}
$$
f_2 \equiv \frac{{\cal B}(c\overline{c} \to \overline{p} + \Lambda_c^+ + 
X^{\prime} + anticharmed~baryon)}
{{\cal B}(c\overline{c} \to \overline{p} + \Lambda_c^+ + X)}
\eqno(10)$$
\end{center}  

\noindent where  
the anticharmed baryon 
could be, e.g. a ${\overline \Lambda}_c$ in which case two baryon 
pairs must exist in the event.

For ${\cal Y}({\overline D}^0|{\overline p})$, 
the yield of events containing a 
${\overline D}^0 \to 
K^+\pi^-$ decay in the 
hemisphere opposite a tag $\overline{p}$, 
one can write:

\begin{center}
$$
{\cal Y}({\overline D}^0|\overline{p}) = \frac{{\cal L} \cdot 
\sigma(e^+e^- \to c\overline{c}) \cdot {\cal B}(c\overline{c} 
\to \overline{p} + \Lambda_c^+ + {\overline D}^0 + X) 
\cdot {\cal B}({\overline D}^0 \to K^+\pi^-) \cdot \epsilon_{\overline{p}} 
\cdot \epsilon_{({\overline D}^0|\overline{p})}}{1-f_3}
\eqno(11)$$
\end{center} 

\noindent and similarly for ${\cal Y}(D^-)$ and ${\cal Y}(D_{s}^-)$, where 
$\epsilon_{({\overline D}^0|\overline{p})}$ is the efficiency for 
reconstructing a ${\overline D}^0 \to K^+\pi^-$ decay in events 
containing a tag anti-proton and where $f_3$ is defined as the fraction of 
($\overline{D}|\overline{p}$) events not containing a $\Lambda_c^+$:

\begin{center}
$$
f_3 \equiv \frac{{\cal B}(c\overline{c} \to \overline{p} + 
\Theta_c^{\prime} + X' + \overline{D})}{{\cal B}
(c\overline{c} \to \overline{p} + X + \overline{D})}
\eqno(12)$$
\end{center}

\noindent in which $\Theta_c^{\prime}$ is a charm+baryon system other than a 
$\Lambda_c^+$. The main contributors to the numerator of this equation are
events like $e^+e^-\to\overline{D}DN\overline{p}X$ and, to a smaller,
negligible extent, events in which a $\Xi_c$ (Sect. \ref{Xic-lambdabar}) is
produced. Note that $f_3$ is closely related
to the previously defined $f_1$ (Eqn. 5); $f_1\approx 1+f_3$.

It then follows that

\begin{center}
$$
{\cal B}(\Lambda_c^+ \to pK^-\pi^+) = 
\frac{\frac{{\cal Y}[\Lambda_c^+\overline{p}] \cdot (1-f_2)}
{\epsilon_{[\Lambda_c^+ \overline{p}]}}}
{\frac{{\cal Y}({\overline D}^0| \overline{p}) \cdot (1-f_3)}
{{\cal B}({\overline D}^0 \to K^+ \pi^-) \cdot 
\epsilon_{({\overline D}^0|\overline{p})}} + 
\frac{{\cal Y}(D^-|\overline{p}) \cdot (1-f_3)}
{{\cal B}(D^- \to K^+ \pi^- \pi^-) \cdot 
\epsilon_{(D^-| \overline{p})}} + 
\frac{{\cal Y}(D^-_{\rm s}| \overline{p}) \cdot 
(1-f_3)}
{{\cal B}(D_{\rm s}^- \to \phi \pi^-) 
\cdot \epsilon_{(Ds^-|\overline{p})}}}
\eqno(13)$$
\end{center}

\noindent where, as before, 
particles contained in [ ] are in the same hemisphere 
with respect to one another and particles contained in ( ) are in opposite 
hemispheres with respect to one another.

The major contributors to $f_2$ are events 
containing $\overline{\Lambda}_c\Lambda_c^+N\overline{p}$. We measure the 
magnitude of this correction by measuring the yield of events 
containing a $\overline{\Lambda}_c$ in the hemisphere opposite a
tag anti-proton.  Our equation for $f_2$ is then

\begin{center}
$$
f_2 = \frac
{
{\cal Y}(\overline{\Lambda}_c|\overline{p})/
\epsilon_{(\overline{\Lambda}_c|\overline{p})}
}
{
{\cal Y}[\Lambda_c^+\overline{p}]/
\epsilon_{[\Lambda_c^+\overline{p}]}
}
\eqno(14)$$
\end{center}

The number of $\overline{D}DN\overline{p}$ events are measured using 
S[$D\overline{p}$] same hemisphere correlations 
(Fig. \ref{fig:pd0}; note that the $DN$
combination here is the major component of
what we previously referred to as ``$\Theta_c'$'');
from these events, we compute $f_3$:

\begin{center}
$$
f_3 = \frac{\frac{{\cal Y}[D^0\overline{p}]}{\epsilon_{[D^0 
\overline{p}]} 
\cdot {\cal B}(D^0 \to K^-\pi^+)} + \frac{{\cal Y}[D^+\overline{p}]}
{\epsilon_{[D^+ \overline{p}]} \cdot 
{\cal B}(D^+ \to K^-\pi^+\pi^+)} 
+ \frac{{\cal Y}[D_{\rm s}^+\overline{p}]}
{\epsilon_{[D_{\rm s}^+\overline{p}]} 
\cdot {\cal B}(D_{\rm s}^+ \to \phi\pi^+)}}{\frac{{\cal Y}
({\overline D}^0|\overline{p})}{\epsilon_{({\overline D}^0|\overline{p})} 
\cdot {\cal B}(D^0 \to K^-\pi^+)} + \frac{{\cal Y}(D^-|\overline{p})}
{\epsilon_{(D^-|\overline{p})} \cdot {\cal B}(D^+ \to K^-\pi^+\pi^+)} 
+ \frac{{\cal Y}(D_{\rm s}^-|\overline{p})}
{\epsilon_{(D_{\rm s}^- |\overline{p})} 
\cdot {\cal B}(D_{\rm s}^+ \to \phi\pi^+)}} 
\eqno(15)$$
\end{center}  

\noindent In this equation, the full expression for 
${\cal Y}[D^0\bar{p}]$, e.g., is

$$ {\cal Y}[D^0\bar{p}] ={\cal
L}\cdot\sigma(e^+e^-\to c\bar{c}) \cdot{\cal B}(c\bar{c}\to
D^0+\bar{p}+\Theta_c+X^\prime)\cdot{\cal B}(D^0\to 
K^-\pi^+)\cdot\epsilon_{\bar{p}}\cdot\epsilon_{[D^0\bar{p}]}.\eqno(15a)  
$$

\subsection{Estimates of $f_2$ and $f_3$}
\subsubsection{$f_3$ and $D{\overline D}N{\overline p}$ backgrounds
{\label{purity2}}}

There are two main contributors to $f_3$: $D\overline{D}N\overline{p}$ events
and fake tag antiprotons.  They were previously discussed in the 
Tag Identification (Sect. III)
and Triple Correlation Sections (Sect. IV), respectively.
Both of these backgrounds inflate the calculated number of
($\overline{D}|\overline{p}$) events (essentially the denominator of
${\cal B}(\Lambda_c^+ \to pK^-\pi^+)$, Eq.(13)) and thus will bias us
towards a low final result if underestimated in the data.
The $D\overline{D}N\overline{p}$ background was found in both data and
Monte Carlo using the plots shown in Figure \ref{fig:pd0} and a similar
one for $D_{\rm s}^-$. 
Monte Carlo simulations indicate that the ${\overline p}$ in
$D\overline{D}N\overline{p}$ is equally likely to appear in the same
hemisphere as the $D$ as in the one opposite the $D$.
Hence the number of events with $D$ and $\overline{p}$ in the same 
hemisphere were simply subtracted from the total number of events in the
denominator of Eqn. (13) for ${\cal B}(\Lambda_c^+ \to
pK^-\pi^+)$.  Numerically, these backgrounds constitute $(17\pm3)$\% and
$(11\pm2)$\% corrections to our initial (${\overline D}|{\overline p}$) sample
in Monte Carlo and data, respectively, as indicated in Table
\ref{tab:event-yields}.  

\subsubsection{$f_3$ and 
$\Lambda_c^+{\overline \Lambda_c}N{\overline p}$ backgrounds}

\begin{figure}[htpb]
\begin{center}
 \begin{picture}(285,85)
 \includegraphics{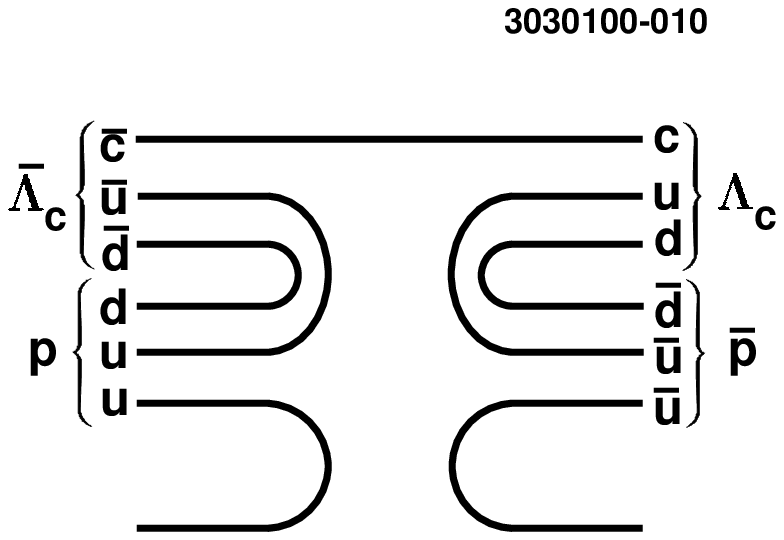}
 \end{picture}
\caption{\label{fig:lamclamcpp_diagram}
\small Possible diagram for producing final states containing four baryons,
including a $\Lambda_c$ opposite a ${\overline\Lambda_c}$.}
\end{center}
\end{figure}

\begin{figure}
\begin{picture}(200,250)
\includegraphics{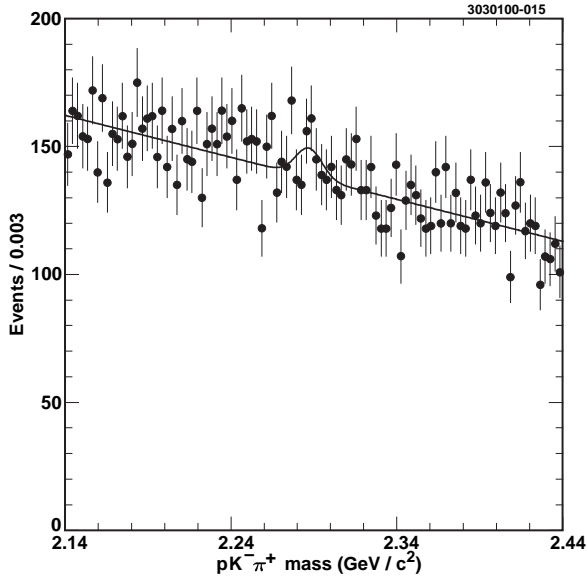}
\end{picture} \\
\caption{\label{fig:plamcplamc}
         \small 
Candidate $\overline{\Lambda}_c$ mass (i.e., $pK^-\pi^+$ mass,
in GeV/$c^2$) for events containing 
a $\overline{p}$ in the opposite hemisphere 
(${\overline p}{\overline\Lambda_c}$).  The 
yield of this plot puts an upper limit on 
$\Lambda_c^+\overline{\Lambda}_cN\overline{p}$ events.}
\end{figure}

There is only one major contributor to $f_2$, namely
$\Lambda_c^+\overline{\Lambda}_cN\overline{p}$ events, as shown in
Figure \ref{fig:lamclamcpp_diagram}.
These events are 
thought to be rare due to the energy needed to create 
the four baryons in such an event. 
However, it is possible that 
$\overline{\Lambda}_c$ production is enhanced when a $\Lambda_c^+$ is produced 
in an event.  In order to estimate this effect we reconstruct 
events containing a $\overline{p}$ opposite a $\overline{\Lambda}_c$ and 
assume a charmed baryon opposite the $\overline{\Lambda}_c$ (see Figure 
\ref{fig:plamcplamc}). 
The effect we see is approximately (7 $\pm$ 3)\% in data.  We therefore 
make an explicit correction of this magnitude ($f_2$).  

\subsection{Particle Reconstruction Efficiency and
Tag Antiproton Fakes}

In deriving ${\cal B}(\Lambda_c^+ \to pK^-\pi^+)$, we assume that 
the Monte Carlo simulation 
accurately reproduces the efficiency for reconstructing
$\Lambda_c^+$'s, that is,
$\epsilon_{\Lambda_c^+(Data)} 
= \epsilon_{\Lambda_c^+(MC)}$.  

Very approximately, the efficiency for reconstructing a 
tag antiproton in the
O(${\overline p}|{\overline D}$)-tagged sample 
should equal the efficiency for reconstructing a tag antiproton in
S[${\overline p}\Lambda_c^+$] events.
However, the latter sample is obviously biased by the high momentum
cut on the $\Lambda_c^+$ ($p_{\Lambda_c^+}>$2.5 GeV/c), which forces the same
hemisphere antiproton tag into a low momentum regime (the kinematic
upper limit on the momentum for a ${\overline p}$ in a signal event to appear
colinear with a
$\Lambda_c^+$ having
$p_{\Lambda_c^+}>$2.5 GeV/c is 
1.65 GeV/$c$). In order to
restrict the $\Lambda_c^+$'s to 
the same momentum interval in our O(${\overline p}|{\overline D}$) sample
(denominator) as those in our S[${\overline p}\Lambda_c^+$] 
sample (numerator), we restrict our calculation of final results to
events in which
tag antiprotons satisfy the requirement $p_{\rm\overline p}<$1.6 GeV/$c$.
This momentum cut therefore helps ensure that the $\Lambda_c^+$ momentum
spectrum in the denominator tag sample 
O(${\overline p}|{\overline D}$) is most similar to the
$\Lambda_c^+$ momentum spectrum in the 
S[${\overline p}\Lambda_c^+$] 
sample, which constitutes the numerator in our double correlation
ratio. 

Below 1.6 GeV/$c$, we must
check that our tag antiprotons in the
O(${\overline p}|{\overline D}$) 
sample have the same momentum spectrum as in our
S[${\overline p}\Lambda_c^+$] sample.
If these subsamples are both drawn from the exact same parent 
([$\Lambda_c^+{\overline p}]|{\overline D}$)
sample, then
we certainly expect this to be the case.
If the tag antiproton momentum spectrum
for our O(${\overline p}|{\overline D}$) sample is the same as 
the tag antiproton momentum spectrum in our 
S[${\overline p}\Lambda_c^+$] sample, then we are
also insensitive to 
any possible variations in the antiproton-finding efficiency as
a function of momentum.

\begin{figure}
\begin{picture}(200,250)
\includegraphics{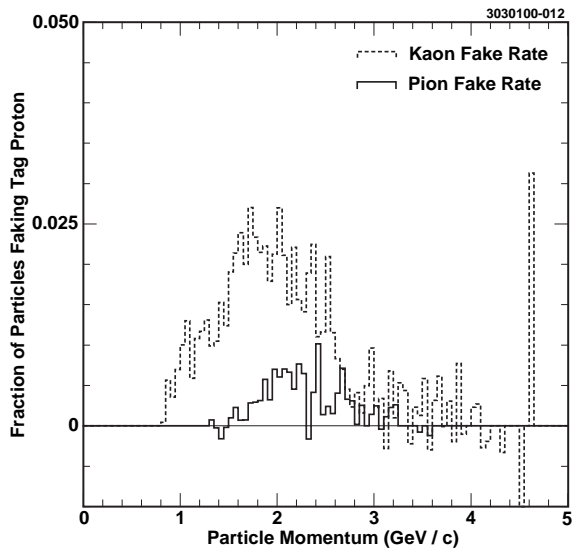}
\end{picture} \\
\caption{\label{fig:fakerates}
         \small The percentage of kaons 
and pions that pass all of our tag proton id 
requirements as a function of momentum.  
The data fake rate was found using kaons and pions from $D^0\to K^-\pi^+$ 
and $\phi\to K^+K^-$
decays as described in the text.}
\end{figure}

Fake antiprotons can also contaminate our candidate antiproton tag sample,
in a momentum-dependent manner.
Figure \ref{fig:fakerates} shows the
likelihood of a kaon track to fake a proton track as a function of momentum,
derived from $\phi\to K^+K^-$ and $D^0\to K^-\pi^+$ events.
Note that the rate at which pions 
fake protons is considerably smaller than the 
rate at which kaons fake protons
(Figure \ref{fig:fakerates})
in the momentum interval of interest ($p<$1.6 GeV/$c$),
since kaons are closer in mass to protons than pions. 
Since pions tend to have random correlations with both $D$-mesons
as well as $\Lambda_c^+$'s, pions faking protons
largely cancel in both numerator
and denominator of Eqn. 13.  This is not necessarily the case for kaons
faking protons.

In $c{\overline c}$ events which do not contain charmed baryons, we expect a
$D$-meson recoiling against the tag ${\overline D}$; the $D$-meson will then
decay into a negatively charged kaon (53$\pm$4)\% of the time if the parent is
a $D^0$ and $(24\pm3)$\% of the time if the parent is a $D^+$\cite{PDG98}.  If
the parent is a $D_{\rm s}$, $K^-$ are produced (13$\pm$13)\% of the
time\cite{PDG98}, hence the population of 
$K^-$ potentially faking ${\overline{p}}$
is enhanced in our ${\overline D}$ tagged sample. 
Unfortunately, we have insufficient statistics to determine the level
of the fake tag background entirely from data, and we must rely on the
Monte Carlo kaon and pion background fractions as a function of momentum to 
quantify antiproton fakes. 

\begin{figure}
\begin{picture}(200,250)
\includegraphics{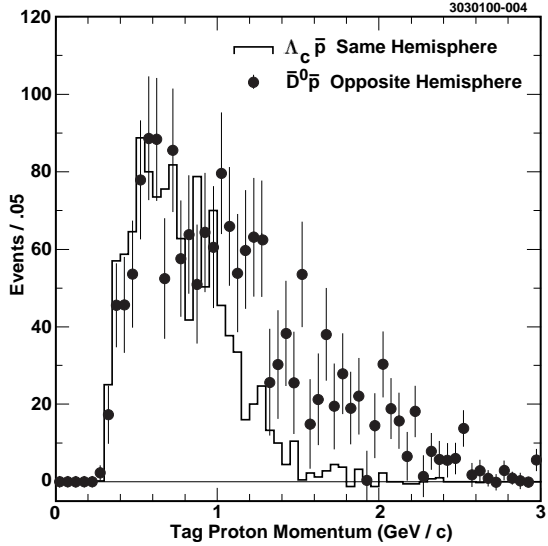}
\end{picture} \\
\caption{\label{fig:beforecorr}
         \small Data tag antiproton 
momentum spectrum in [$\Lambda_c^+\overline{p}$] 
same hemisphere events (solid histogram) and 
$\overline{D}$ opposite $\overline{p}$ 
O($\overline{D}|\overline{p}$) events (points) in data, 
after a sideband subtraction on the $\Lambda_c^+$ or $\overline{D}$ mass, 
prior to subtracting tag anti-proton fakes.  
Notice the excess of high momentum tag antiprotons in the 
O$(\overline{D}|\overline{p})$ sample.} 
\end{figure}

\begin{figure}
\begin{picture}(180,220)
\includegraphics{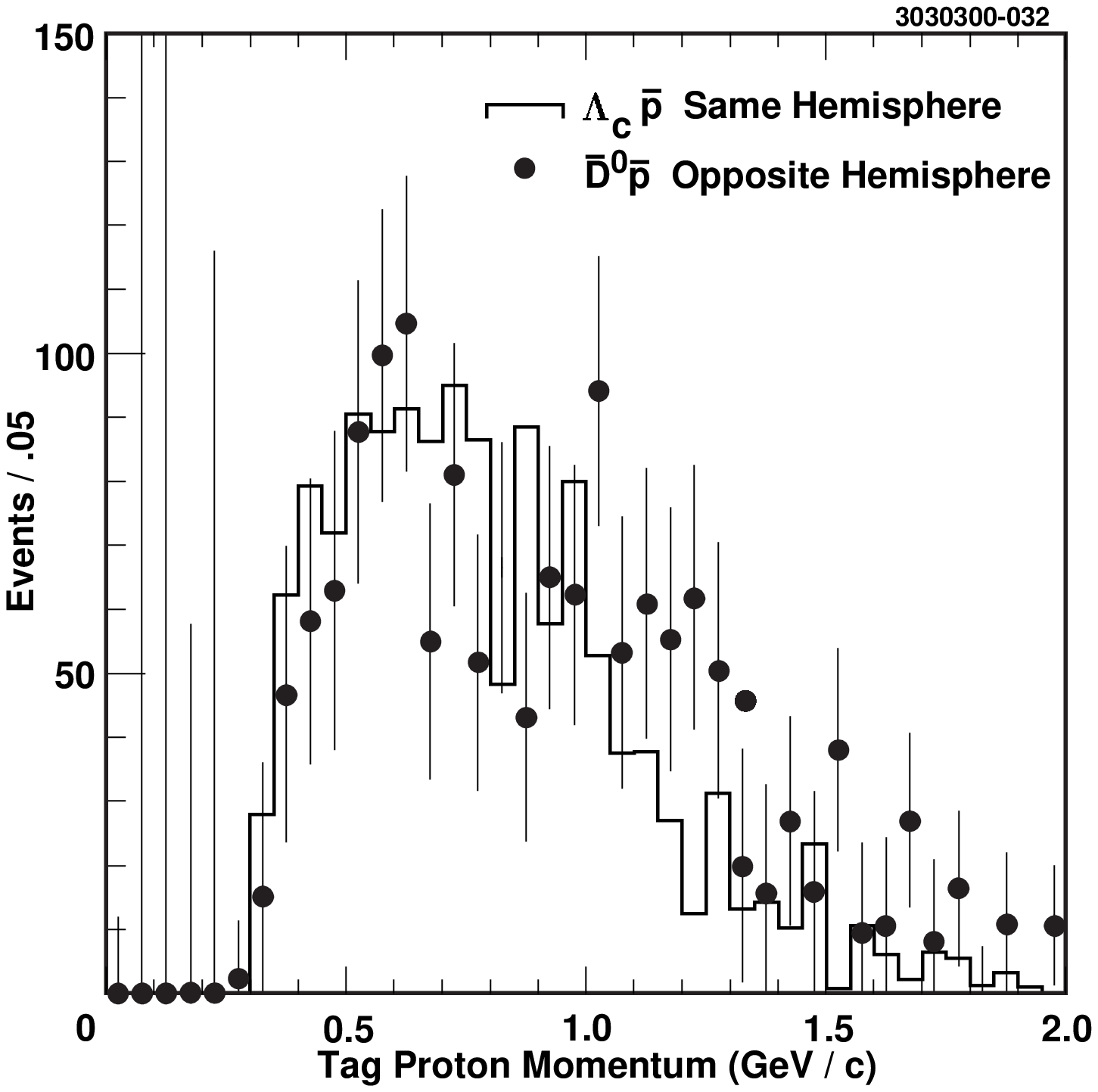}
\end{picture} 
\begin{picture}(180,220)
\includegraphics{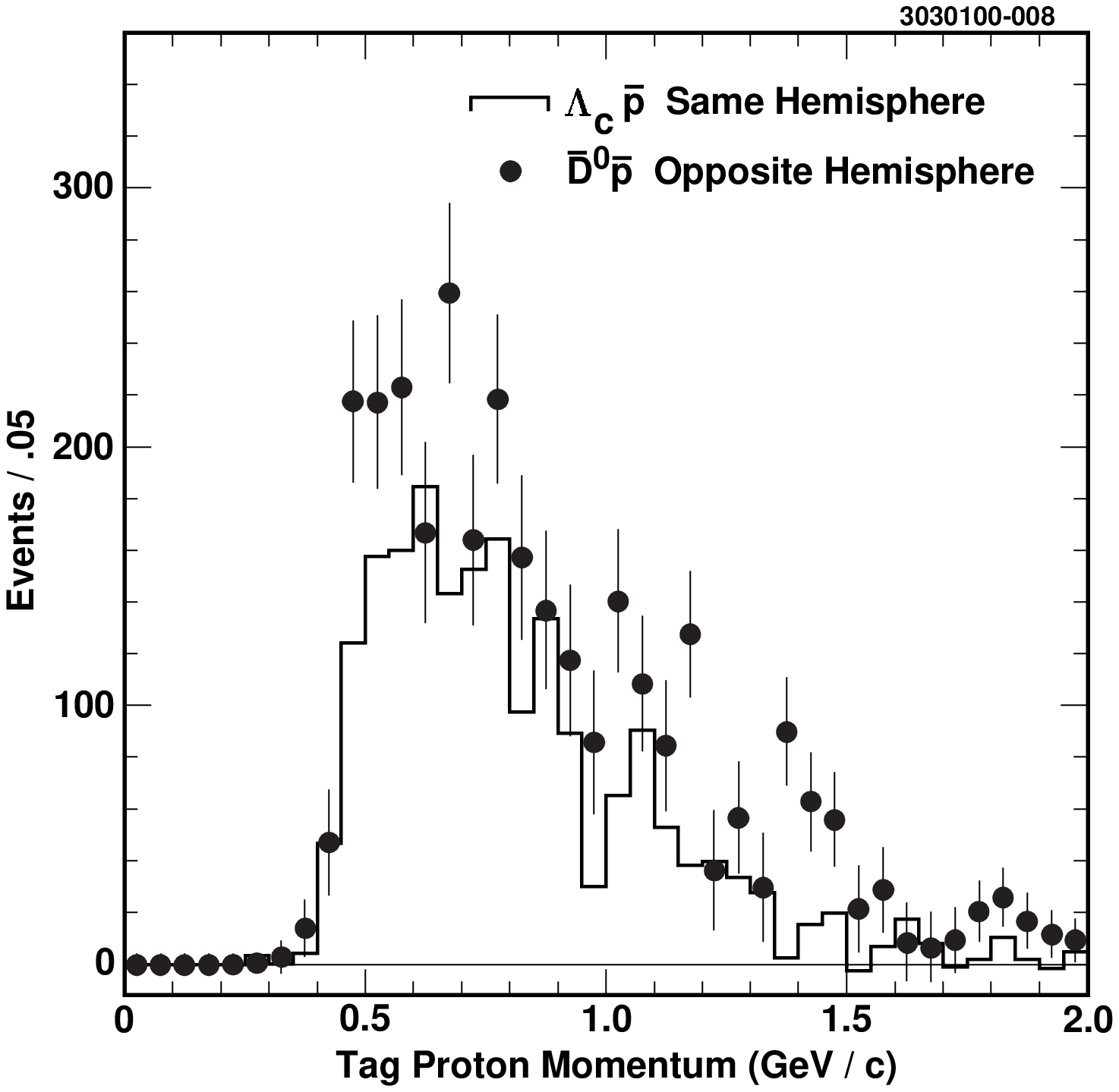}
\end{picture} \\
\caption{\label{fig:fakecorr}
         \small Left:
Previous plot after background subtractions.
Tag anti-proton momentum in $\Lambda_c^+\overline{p}$ 
same hemisphere events ([$\Lambda_c^+{\overline p}$], solid histogram) and 
$\overline{D}$ opposite $\overline{p}$ events (O($\overline{D}|{\overline p}$),
points) in data
after a sideband subtraction on the $\Lambda_c^+$ or $\overline{D}$ mass, 
and
after subtracting tag antiproton fakes. Right: Corresponding
Monte Carlo spectra, after similar subtractions,
for comparison.}  
\end{figure}

We thus use the following
procedure to determine the contribution of fakes to our tag antiproton
sample and then
extract our final branching fraction:
\begin{enumerate}
\item We plot the tag anti-proton momentum spectrum, separately for our
O(${\overline p}|{\overline D}$) and S[${\overline p}\Lambda_c^+$] 
samples prior to any corrections
(Figure \ref{fig:beforecorr}). Since there is some background under
the $\Lambda_c^+$ and ${\overline D}$ mass distributions,
a sideband subtraction must be performed to remove background 
$(pK^-\pi^+)$-${\overline p}$ correlations in the case of the 
S[${\overline p}\Lambda_c^+$] sample, with a similar sideband subtraction
for the O(${\overline p}|{\overline D}$) sample.
The scaled antiproton momentum spectrum opposite 
$K^+\pi^-$ 
invariant mass
combinations in the ${\overline D}^0$ sidebands 
($0.03<|m_{K^+\pi^-}-m_{\overline D^0}|<0.1$ GeV) is therefore
subtracted from the antiproton momentum spectrum opposite 
$K^+\pi^-$ 
invariant mass
combinations in the ${\overline D}^0$ signal region
($|m_{K^+\pi^-}-m_{\overline D^0}|<0.025$ GeV).
We note in Figure \ref{fig:beforecorr} a large excess above
the $p<1.6$ GeV/c kinematic limit, which we attribute,
in part, to backgrounds from 
$D{\overline D}N{\overline p}$ and kaons producing fake antiproton tags.
\item 
We now remove
the contribution from non ${\overline p}{\overline D}\Lambda_c^+$ 
events (in both
data and Monte Carlo) to each of our
O(${\overline p}|{\overline D}$) and S[${\overline p}\Lambda_c^+$] 
samples (separately).
\begin{enumerate}
\item We first subtract fake antiprotons using the measured kaon/pion
fake rates as a function of momentum, multiplied by the kaon and pion
production rates 
as a function of momentum. The per track fake
rates are determined directly from data, as described previously.
For the production momentum spectra, we rely on
Monte Carlo simulations, which are based on the
Particle Data Group $D\to K^-X$ exclusive branching fractions
and inclusive rates\cite{PDG98}. 
\item 
We additionally subtract contributions due to
$D{\overline D}N{\overline p}$ 
and $\Lambda_c^+{\overline p}N{\overline\Theta_c}$ from the
fake-subtracted plot, using data for both of these estimates.
These backgrounds are estimated from the yields in
Figs. \ref{fig:pd0} and \ref{fig:plamcplamc}, respectively.
The sideband-subtracted antiproton momentum spectra in our
S[$p{\overline D}$] and O(${\overline p}|{\overline\Lambda_c^-}$)
data samples are themselves
directly subtracted from the signal
O(${\overline p}|{\overline D}$) and S[${\overline p}{\Lambda_c^+}$]
antiproton momentum spectra.
In doing so, we have 
removed backgrounds from
$D{\overline D}N{\overline p}$ 
and $\Lambda_c^+{\overline p}N{\overline\Theta_c}$.
\end{enumerate}
After subtracting these backgrounds, we note 
improved agreement between the data tag antiproton spectrum for the 
O($\overline{p}|\overline{D}$) and S[$\overline{p}\Lambda_c^+$] samples.
(Fig. \ref{fig:fakecorr}).
\item 
After performing the above subtractions, 
we extract ${\cal B}(\Lambda_c^+\to pK^-\pi^+)$,
restricting ourselves to the interval where the Monte Carlo and data show good
agreement for the ${\overline p}$ momentum spectra
(0.6--1.6 GeV/$c$; as already mentioned,
the upper momentum cut coincides with the kinematically allowed 
maximum
momentum for our tag antiprotons 
given the minimum momentum requirement on
the $\Lambda_c^+$).
\end{enumerate}
We take a combination of 
the magnitude of the fake subtraction and the
spread in the derived values of ${\cal B}(\Lambda_c^+\to pK^-\pi^+)$ when
we vary the limits of our tag antiproton momentum acceptance, as an
estimate of the
systematic error inherent
in this procedure ($\sim$15\%, Table II).

\subsection{Checks of the ${\overline p}$ momentum spectrum}

We have conducted a check of the double correlation analysis
by using a sample
of events which have the tag antiproton in the opposite (same) hemisphere
of (as)
the $\Lambda_c^+$ (${\overline D}$), i.e., opposite to the
correlation exploited in our standard analysis. 
Following the above notation, we denote these
events as ([${\overline p}{\overline D}]|\Lambda_c^+$) events, and the
subsample of those 
events which 
constitute the denominator and
numerator in our double correlation sample as
S[${\overline p}{\overline D}$] and 
O(${\overline p}|\Lambda_c^+$), 
respectively.
According to JETSET 7.3 simulations, approximately half
of all ${\overline p}{\overline D}$$\Lambda_c^+$ 
events will have the antiproton in the same
hemisphere as the $\Lambda_c^+$ (corresponding to
our ``standard'' analysis), with the other half having the antiproton in
the opposite hemisphere (see Fig.
\ref{fig:hemispheres}).  We do not use these hemisphere/sign correlations in
computing ${\cal B}(\Lambda_c^+ \to pK^-\pi^+)$ for two main reasons - first,
the level of $D{\overline D}N{\overline p}$ is much more difficult to
determine than for the standard O(${\overline p}|{\overline D}$) sample, and
second, the S[${\overline p}\Lambda_c^+$] sample is very susceptible to
$\overline{\Lambda}_c\Lambda_c^+$ events in which there are no charmed mesons
produced - in such a case, the tag antiproton can be a direct decay product
of the $\overline{\Lambda}_c$.
If we nevertheless trust the Monte Carlo to reproduce all backgrounds and
efficiencies for this S[${\overline p}{\overline D}$]-tagged 
data sample, and calibrate the observed
yields in data to Monte Carlo simulations as above in the standard
double correlation analysis, we obtain a central
value for ${\cal B}(\Lambda_c^+\to pK^-\pi^+)$
which differs by $\sim$12\% from the standard analysis (see Table II).

\begin{figure}
\begin{picture}(200,250)
\includegraphics{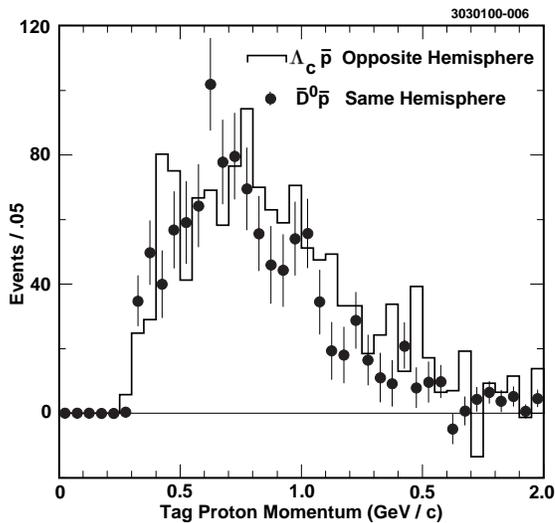}
\end{picture} \\
\caption{\label{fig:fakealt}
         \small Cross-check results.
Tag antiproton momentum in O($\Lambda_c^+|\overline{p}$)
opposite hemisphere events (solid histogram) and 
S[$\overline{D} \overline{p}$] same hemisphere events (points) in data, 
after a sideband subtraction on the $\Lambda_c^+$ or $\overline{D}^0$ mass.  
This sample should be less susceptible to tag proton fakes, since the 
sign correlation is not correct for kaons coming from 
semileptonic charm decay to
fake antiprotons, as 
was the case for our $\overline{D}$ opposite $\overline{p}$ event sample.  
However, we do not use these hemisphere/sign correlations in computing 
${\cal B}(\Lambda_c^+ \to pK^-\pi^+)$ due to the difficulty in 
estimating the
$D\overline{D}N\overline{p}$ and $\Lambda_c^+{\overline\Lambda_c}$ 
backgrounds in this event sample.}  
\end{figure}

This ([${\overline p}{\overline D}]|\Lambda_c^+$) 
sample is much less
susceptible to backgrounds from $K^-$ which fake ${\overline p}$ because
the $D$, which would be the putative source of these fakes, is now 
fully reconstructed.
The requirement that the tag antiproton now be found in the ${\overline
D}$ hemisphere rather than the $\Lambda_c^+$ hemisphere biases the tag
antiproton momentum spectrum in a different way than in the ``standard''
analysis.  
We can thus use these ($\Lambda_c^+|[{\overline p}{\overline D}$]) 
events to qualitatively check our tag ${\overline p}$
momentum spectra in the ``standard'' 
([${\overline p}\Lambda_c^+]|{\overline D}$)
analysis, after kaon fake subtraction in the 
standard analysis.
The momentum spectrum for tag antiprotons in
our cross-check 
($\Lambda_c^+|[{\overline p}{\overline D}$]) sample is shown in
Figure \ref{fig:fakealt}. 
The antiproton momentum spectrum in the
($\Lambda_c^+|[{\overline p}{\overline D}$])
sample is qualitatively similar to that in the
standard
([$\Lambda_c^+{\overline p}]|{\overline D}$) analysis, after background
corrections.

\section{Results}
Our results, showing the
yields ${\cal Y}$, efficiencies $\epsilon$, and
backgrounds, in both data and Monte Carlo, are tabulated in
Table \ref{tab:event-yields}.
The weighted average of the three techniques corresponds to
${\cal B}(\Lambda_c^+ \to pK^-\pi^+)=(5.0\pm0.5)$\% (statistical error only).

\begin{table}[htpb]
\caption{Event yields for signal and backgrounds, in data and Monte 
Carlo. As before, same hemisphere correlations are
designated with brackets [~], and opposite hemisphere
correlations are designated with parenthesis (~). Background
yields which
are subtracted from the numerator or denominator
are indicated with a minus sign.}\label{tab:event-yields}
\begin{small}
\begin{center}
\begin{tabular}{c|cc}      \hline   \hline
Double Correlations & MC& Data \\ \hline 
${\cal Y}[{ \Lambda_c^+ \overline{p}}]$ (Numerator) & 
1656$\pm$65 & 1093$\pm$47 \\
${\cal Y}({{\overline D}^0}|{\overline{p}})$ (Denominator)
& 2725$\pm$84 & 1369$\pm$55 \\ 
${\cal Y}({ D^-}|{\overline p})$ (Denominator) 
& 1501$\pm$113 & 963$\pm$71 \\ 
${\cal Y}({ D_{\rm s}^-}|{ \overline p})$ (Denominator) 
& 111$\pm$19 & 51$\pm$11  \\ 
$\epsilon_{\Lambda_c^+}/\epsilon_{D^0}/\epsilon_{D^-}/\epsilon_{D_{\rm s}^-}$ 
& 26.5\%/43.7\%/32.2\%/14.5\% \\
${\cal Y}({\overline{\Lambda}_c}|{\overline p})$ (Bkgnd to Num.) 
& --84$\pm$49 & --75$\pm$39 \\ 
${\cal Y}[{ D^0\overline p}]$ (Bkgnd to Den.) & --268$\pm$40 & 
--68$\pm$23 \\ 
${\cal Y}[{ D^+\overline p}]$ (Bkgnd to Den.) & --417$\pm$67 & 
--152$\pm$39 \\ 
${\cal Y}[{ D_{\rm s}^+\overline p}]$ (Bkgnd to Den.) & --26$\pm$11 & 
--1$\pm$6  \\ 
fake ${\overline p}$ in $[D{\overline p}]$ evts.
(Bkgnd. to Den.) & --272$\pm$31 & --298$\pm$36 \\ \hline
$f_2$ & ($5.1\pm2.9$)\%  & ($6.9\pm2.2$)\%  \\ \hline
$f_3$ ($\approx f_1-1$) & ($17.5\pm3.4$)\%  & (10.6$\pm$2.4)\%  \\ \hline
${\cal B}(\Lambda_c^+ \to pK^-\pi^+)$ & 4.3\% (input) 
& (4.9$\pm$0.5)\% \\ \hline \hline
(${ \pi^-_{\rm soft}}|{\overline p})$ Triple Correlation & \\ \hline
${\cal Y}$(${ \pi^-_{\rm soft}}|{\overline p})$ (Denominator) & 
34222$\pm$1092 &
14553$\pm$485 \\
fake ${\overline p}$ in $({ \pi^-_{\rm soft}}|{\overline p})$ &
--3318$\pm$310 & --$1867\pm261$ \\
${\cal Y}([{ \Lambda_c^+\overline p}]|{ \pi_{\rm soft}^-})$ (Numerator) &
202.8$\pm$27.8 & 101.6$\pm$20.6 \\
${\cal B}(\Lambda_c^+ \to pK^-\pi^+)$ & 4.3\% (input) 
& (5.2$\pm$1.3)\% \\ \hline \hline
 $({ e^-}|{\overline p})$ Triple Correlation & \\ \hline
${\cal Y}({ e^-}|{\overline p})$ (Denominator) 
& 4178$\pm$65 & 1739$\pm$47 \\
fake ${\overline p}$ + fake $e^-$ & --382$\pm$39 & --$272\pm 41$ \\
${\cal Y}([{\Lambda_c^+\overline p}]|{ e^-})$ (Numerator) &
20.1$\pm$5.2 & 10.3$\pm$3.8 \\ 
${\cal B}(\Lambda_c^+ \to pK^-\pi^+)$ & 4.3\% (input)
& (5.6$\pm$2.5)\% \\ \hline \hline
\end{tabular}
\end{center}
\end{small}
\end{table}

\newpage
\section{Summary of Systematic Uncertainties}

We have already discussed many of the systematic errors and their 
assessment in previous sections.
Table \ref{tab:systematics} lists the systematic 
errors evaluated for the 
three methods of extracting ${\cal B}(\Lambda_c^+ \to pK^-\pi^+)$. As discussed
previously, the largest systematic error is due to uncertainties in the
tagging efficiency and spectrum. This includes
possible backgrounds to the antiproton tags, and the difference
between the ${\overline p}$ 
momentum spectra in S$[\Lambda_c{\overline p}]$ and 
O$({\overline D}|{\overline p})$ events.
Uncertainties in backgrounds and tagging efficiencies are
assessed, in part, 
by varying the tag antiproton
momentum interval over which our final result is extracted by
$\pm$300 MeV/c in either direction from the default value.
The error (``Event Selection/MC Mismodeling'') is evaluated by
varying the event selection criteria for both data and Monte Carlo and
determining the variation in the calculated final result. 
This error also includes the discrepancy between the
central value we quote and
the result obtained from the cross-check in which the antiproton is
identified in the same hemisphere as the charm tag.
It also includes the variation in the final result obtained using different 
versions of charged track reconstruction software, comparing the
internal consistency of different data subsamples, and different versions
of the Monte Carlo event generator and detector simulation. 
 
\begin{table}[htpb]
\caption{Summary of systematic errors assessed in measurement
of ${\cal B}(\Lambda_c^+\to pK^-\pi^+)$.}\label{tab:systematics}
{\small
\begin{center}
\begin{tabular}{|c||c||c||c|}           \hline
& $\pi_{\rm soft}$ tag & Electron tag & Double Corr. \\ \hline
Tag proton id/spectrum & 15\% & 15\% & 15\% \\ \hline
Event Selection/MC Mismodeling & 15\% & 14\% & 12\% \\ \hline \hline
${ D}{\overline D}{ N}{\overline p}$ background events 
& 8\% & 8\% & 8\% \\ \hline
$\Lambda_c^+/D$ momentum spectra & 8\% & 8\% & 8\% \\ \hline
$\Lambda_c^+/D$ id cuts & 5\%& 5\% & 5\% \\ \hline
$\Lambda_c^+$ mass fit/sideband subtraction & 6\% & 10\%  & 1\% \\ \hline
Contamination from $\Sigma_c^0$, $\Lambda_{cJ}$ & 6\% & - & - \\ \hline
tag electron fakes & - & 5\% & - \\ \hline
${\Lambda_c^+}{\overline\Lambda_c}{ N}{\overline p}$ events 
& - & - & 4\% \\ \hline
${\cal B}(D^0 \to K^-\pi^+)$ & - & - & 2\%\\ \hline
${\cal B}(D^+ \to K^-\pi^+\pi^+)$ & - & - & 3\% \\ \hline 
Beam-wall/Beam-gas contamination & 3.5\% & 3.5\% & 3.5\% \\ \hline
Contamination of $\Xi_c$, $\Omega_c$ events & 3\%  & 3\% & 3\% \\ \hline
Hemisphere Correlation & 2\% & 2\% & 2\% \\ \hline \hline
Total & 28\%  & 25\%  & 24\% \\ \hline
\end{tabular}
\end{center}}
\end{table}

\section{Discussion and Conclusions}
Employing new techniques of baryon-charmed particle correlations in
$e^+e^-\to c{\overline c}$ annihilations at a center of mass
energy $\sqrt{s}\sim$10.55 GeV,
we measure
${\cal B}(\Lambda_c^+\to pK^-\pi^+)=(5.0\pm0.5\pm1.2)$\%.
At present, this technique is limited by our understanding of the
non-signal backgrounds (most notably, $D{\overline D}N{\overline p}$
backgrounds); presumably, more data would allow a greater 
understanding of those backgrounds.
Our result is
consistent with the determination of
${\cal B}(\Lambda_c^+\to pK^-\pi^+)$=7$\pm$2\%
suggested 
by Dunietz\cite{Isi98}, based on the measured ratio
for ${\cal B}(\Lambda_c^+\to\Lambda Xl\nu)/{\cal B}(\Lambda_c^+\to pK^-\pi^+)$ 
and 
assuming that the semileptonic charmed baryon width is the same
as the semileptonic charmed meson width.
It is also consistent with the value of $(5.0\pm1.3)$\% derived by the
Particle Data Group\cite{PDG98}.
We now
discuss the implications of this result and its consistency with related
measurements.

The product branching fraction:
${\cal B}(B\to(\Lambda_c^+X~or~{\overline{\Lambda}_c}X))
\cdot{\cal B}(\Lambda_c^+\to pK^-\pi^+)$ 
can be directly
determined by simply measuring the efficiency-corrected $\Lambda_c^+\to 
pK^-\pi^+$ 
yield in $B{\overline B}$ 
events. An unpublished CLEO result finds
a value of
${\cal B}((B+{\overline B})\to\Lambda_c^+)\cdot(\Lambda_c^+\to pK^-\pi^+)=
(1.81\pm0.22\pm0.24)\times 10^{-3}$\cite{Zoeller} for this product
branching fraction. Given that, 
${\cal B}(\Lambda_c^+\to pK^-\pi^+)=0.05$
implies that
${\cal B}(B\to(\Lambda_c^+~or~{\overline\Lambda_c}))\sim$3.6\%. 
This can be compared
to the Particle Data Group
value of ${\cal B}(B\to p~or~{\overline p})\sim$8.0\%\cite{PDG98}. 
Our result therefore implies that $B\to baryons$ may be occurring at
a substantial rate
through modes such as ${\overline B}\to 
DN{\overline N}$X\cite{Dunietz-BDNNbar}, 
${\overline B}\to\Xi_c{\overline Y}$X, or 
${\overline B}\to\Xi_c{\overline\Lambda_c}$.
CLEO has recently published evidence for the latter modes\cite{Horst-paper}.

We can also place bounds on the $\Lambda_c^+\to pK^-\pi^+$ branching fraction
by using the measured CLEO $e^+e^-\to$hadrons cross-section, assuming that
the $c{\overline c}$ fraction is 40\% of the total hadronic cross-section.
CLEO has measured 
${\cal B}(\Lambda_c^+\to pK^-\pi^+)\cdot\sigma(e^+e^-\to 
(\Lambda_c^++{\overline\Lambda_c}))=10\pm1$ pb. That measurement simply
determines the total yield of either $\Lambda_c^+$ or
${\overline\Lambda_c}$ in $e^+e^-$ annihilations; i.e., it determines the
sum of $c\to\Lambda_c^+$ plus ${\overline c}\to{\overline\Lambda_c}$.
Our value of ${\cal B}(\Lambda_c^+\to pK^-\pi^+)$=0.05 implies that
$\sigma(e^+e^-\to (\Lambda_c^++{\overline\Lambda_c}))=$200 pb. 
Using the recent CLEO measurement of 
$R\equiv{\sigma(e^+e^-\to q{\overline q})\over
\sigma(e^+e^-\to\mu^+\mu^-)}$\cite{CLEOR}, which corresponds to
a value of $\sigma(e^+e^-\to q{\overline q})\sim$3.3 nb, 
and using the JETSET value of $c\to\Lambda_c^+\sim$0.07, we
have:
$\sigma(e^+e^-\to q{\overline q})\times
{{c\overline c \over q\overline q}}\times
(c\to\Lambda_c^++{\overline c}\to{\overline\Lambda_c})$=
3300 pb$\times 0.4 \times 0.07\times 2$=185 pb, in good agreement with
our measurement above.

Finally, since the presently tabulated exclusive $\Lambda_c^+$ decays
are all normalized to ${\cal B}(\Lambda_c^+\to pK^-\pi^+)$, we
conclude that $\sim$50\% of the $\Lambda_c^+$ width is unaccounted for.
Since the $\Lambda_c^+$ lifetime is only $\sim$40\% of the 
$D^0/D_{\rm s}$ lifetime, it has long been realized that diagrams such as 
exchange diagrams, and/or final states
including neutrons, are
likely to be large contributors to $\Lambda_c^+$ decay and
may produce final states different than the `usual' states expected
from simple $\Lambda_c^+\to\Lambda W_{\rm external}$ diagrams. Measurement of
such decays await additional data and analysis.

\acknowledgments
\label{sec:acknowledgments}

We gratefully acknowledge the effort of the CESR staff in providing us with
excellent luminosity and running conditions.
J.R. Patterson and I.P.J. Shipsey thank the NYI program of the NSF, 
M. Selen thanks the PFF program of the NSF, 
M. Selen and H. Yamamoto thank the OJI program of DOE, 
J.R. Patterson, K. Honscheid, M. Selen and V. Sharma 
thank the A.P. Sloan Foundation, 
M. Selen and V. Sharma thank the Research Corporation, 
F. Blanc thanks the Swiss National Science Foundation, 
and H. Schwarthoff and E. von Toerne thank 
the Alexander von Humboldt Stiftung for support.  
This work was supported by the National Science Foundation, the
U.S. Department of Energy, and the Natural Sciences and Engineering Research 
Council of Canada.

\end{document}